\documentclass[iop,twocolappendix]{emulateapj}

\usepackage{natbib}
\usepackage{amsmath}
\usepackage{multirow}
\usepackage[]{threeparttable}
\usepackage{verbatim}
\usepackage{color}

\usepackage{hyperref}

\shorttitle{2LAC-blazar contribution to TeV-PeV neutrinos}
\shortauthors{M.~G.~Aartsen et al.} 

\begin{document}


\title{The contribution of Fermi-2LAC blazars to the diffuse TeV-PeV neutrino flux}

\author{
IceCube Collaboration:
M.~G.~Aartsen\altaffilmark{1},
K.~Abraham\altaffilmark{2},
M.~Ackermann\altaffilmark{3},
J.~Adams\altaffilmark{4},
J.~A.~Aguilar\altaffilmark{5},
M.~Ahlers\altaffilmark{6},
M.~Ahrens\altaffilmark{7},
D.~Altmann\altaffilmark{8},
K.~Andeen\altaffilmark{9},
T.~Anderson\altaffilmark{10}
% ...
}

\altaffiltext{1}{Department of Physics, University of Adelaide, Adelaide, 5005, Australia}
\altaffiltext{2}{Physik-department, Technische Universit\"at M\"unchen, D-85748 Garching, Germany}
\altaffiltext{3}{DESY, D-15735 Zeuthen, Germany}
\altaffiltext{4}{Dept.~of Physics and Astronomy, University of Canterbury, Private Bag 4800, Christchurch, New Zealand}
\altaffiltext{5}{Universit\'e Libre de Bruxelles, Science Faculty CP230, B-1050 Brussels, Belgium}
\altaffiltext{6}{Dept.~of Physics and Wisconsin IceCube Particle Astrophysics Center, University of Wisconsin, Madison, WI 53706, USA}
\altaffiltext{7}{Oskar Klein Centre and Dept.~of Physics, Stockholm University, SE-10691 Stockholm, Sweden}
\altaffiltext{8}{Erlangen Centre for Astroparticle Physics, Friedrich-Alexander-Universit\"at Erlangen-N\"urnberg, D-91058 Erlangen, Germany}
\altaffiltext{9}{Department of Physics, Marquette University, Milwaukee, WI, 53201, USA}
\altaffiltext{10}{Dept.~of Physics, Pennsylvania State University, University Park, PA 16802, USA}
% ...

\begin{abstract}

The recent discovery of a diffuse cosmic neutrino flux extending up to PeV energies raises the question of which astrophysical sources generate this signal. One class of extragalactic sources which may produce such high-energy neutrinos are blazars. We present a likelihood analysis searching for cumulative neutrino emission from blazars in the 2nd Fermi-LAT AGN catalogue (2LAC) using an IceCube neutrino dataset 2009-12 which was optimised for the detection of individual sources. In contrast to previous searches with IceCube, the populations investigated contain up to hundreds of sources, the largest one being the entire blazar sample in the 2LAC catalogue. No significant excess is observed and upper limits for the cumulative flux from these populations are obtained. These constrain the maximum contribution of the 2LAC blazars to the observed astrophysical neutrino flux to be $27 \%$ or less between around 10 TeV and 2 PeV, assuming equipartition of flavours at Earth and a single power-law spectrum with a spectral index of $-2.5$. We can still exclude that the 2LAC blazars (and sub-populations) emit more than $50 \%$ of the observed neutrinos up to a spectral index as hard as $-2.2$ in the same energy range. Our result takes into account that the neutrino source count distribution is unknown, and it does not assume strict proportionality of the neutrino flux to the measured 2LAC $\gamma$-ray signal for each source. Additionally, we constrain recent models for neutrino emission by blazars.

\end{abstract}

\cleardoublepage

\section{Introduction}
\label{sec:introduction}

The initial discovery of the astrophysical neutrino flux around PeV energies \citep{Aartsen2013} a few years ago marked the beginning of high-energy neutrino astronomy. Since then, its properties have been measured with increasing accuracy \citep{Aartsen2015}. The most recent results indicate a soft spectrum with a spectral index of $-2.5 \pm 0.1$ between around 10 TeV and 2 PeV with no significant deviation from an equal flavor composition at Earth \citep{Aartsen2015a}. 
The neutrino signal has been found to be compatible with an isotropic distribution on the sky. This apparent isotropy suggests that a significant fraction of the observed neutrinos is of extragalactic origin, a result which is also supported by \citet{Ahlers2015}. However, there are also indications for a 3-$\sigma$ anisotropy \citep{Neronov2016} if low-energy events ($<100$ TeV) are omitted. Further data are required to settle this issue.


Potential extragalactic source candidates 
are Active Galactic nuclei (\textbf{AGN}), where both radio-quiet \citep{Stecker1991} and radio-loud \citep{Mannheim1995} objects have been considered for neutrino production since many years. 
Blazars, a subset of radio-loud active galactic nuclei with relativistic jets pointing towards Earth \citep{Urry1995}, are investigated in this paper. They are commonly classified based on the properties of the spectral energy distribution (\textbf{SED}) of their electromagnetic emission. The blazar SED features two distinctive peaks: a low-energy peak between infrared and X-ray energies, 
attributed to synchrotron emission of energetic electrons, and a high-energy peak at $\gamma$-ray energies, which can be explained by several and possibly competing interaction and radiation processes of high-energy electrons and high-energy nuclei \citep{Boettcher2013}.
Several works suggest that blazar SEDs follow a sequence \citep{Fossati1998, Bottcher2002, Cavaliere2002,Meyer2011}, in which the peak energy of the synchrotron emission spectrum decreases with increasing blazar luminosity. Accordingly, blazars can be classified into low synchrotron peak (\textbf{LSP}), intermediate synchrotron peak (\textbf{ISP}) and high synchrotron peak (\textbf{HSP}) objects\footnote{This scheme is a generalization of the XBL/RBL classification of BL\,Lac objects introduced by \citet{Padovani1995}.}, a classification scheme introduced in \citet{Abdo2010} which we use throughout this work. A second classifier is based on the prominence of emission lines in the SED over the non-thermal continuum emission of the jet. Flat Spectrum Radio Quasars (\textbf{FSRQ}s) show Doppler-broadened optical emission lines \citep{Stickel1991}, while in so called BL\,Lac objects emission lines are hidden under a strong continuum emission.

Many calculations of high-energy neutrino emission from the jets of blazars can be found in the literature. Neutrinos could be produced via charged pion decay in interactions of high-energy protons 
with gas (pp-interactions) in the jets \citep{Schuster2002} or in interactions of protons with internal \citep{Mannheim1995} or 
external \citep{Atoyan2001} photon fields (p$\gamma$-interactions).
Early models for the neutrino emission from blazars made no explicit distinction based on the blazar class. Some of these have already been explicitly excluded at $90 \%$ C.L. by past diffuse neutrino flux measurements \citep{Abbasi2011, Aartsen2014a}, for example the combined pp+p$\gamma$ predictions in \citet{Mannheim1995}.
More recent publications, on the other hand, differentiate between specific classes of blazars and are largely not constrained by experiment, yet. The neutrino production of BL\,Lac objects is modeled e.g. 
in \citet{Muecke2003,Tavecchio2014,Padovani2015} while neutrino production of FSRQs is calculated e.g. in \citet{Becker2005,Murase2014}. The models by \citet{Tavecchio2014} and \citet{Padovani2015} were in particular constructed to explain parts or all of the astrophysical neutrino flux. 
With the analysis presented here, we are able to test large parts of the parameter space of many of these models for the first time.
We do not consider theoretical calculations from the literature for individual objects, since these are not directly comparable to our results. 

The neutrinos predicted by most models are produced in charged pion decays which come with an associated flux of $\gamma$-rays from neutral pion decays. 
Even if the hadronic fraction is sub-dominant, one could on average expect a higher neutrino luminosity for a higher observed $\gamma$-luminosity \citep{Murase2014}. 
On a source-by-source basis, however, variations in the exact $\nu/\gamma$ correlation are likely. One strategy to cope with this uncertainty, which we follow in this paper, is to analyze large samples of objects and thereby to investigate average properties. We use the Fermi-LAT 2LAC catalogue\footnote{The successor catalogue 3LAC \citep{Ackermann2015} was not yet published when this analysis was carried out. For the $\gamma$-weighting scheme (see section \ref{sec:gamma_weighting}), the results are expected to be nearly identical. The 2LAC sample already resolves the majority of the GeV-blazar flux and the brightest blazars are also expected to be bright in the 3LAC catalogue in the quasi-steady approximation.} \citep{Ackermann2011}  to define search positions for our analysis  (see section \ref{section:blazar_populations}). The blazars in the 2LAC catalogue comprise the majority ($ \approx 70 \%$) of the total $\gamma$-ray flux emitted from all GeV-blazars in the observable universe between $100 \ \mathrm{MeV}$ and $100 \ \mathrm{GeV}$ (see appendix \ref{appendix:correction_factor}). Compared to other Fermi catalogues starting at higher energies, 1FHL \citep{Ackermann2013} or 2FHL \citep{Ackermann2016}, the 2LAC contains more than twice the number of blazars.
The goal is to look for a cumulative neutrino flux excess from all 862 2LAC blazars or from specifically selected sub-populations using muon-track data with an angular resolution of about a degree in an unbinned maximum-likelihood stacking approach. We use two different "weighting schemes" (see section \ref{section:weighting_schemes}) to define the probability density functions (\textbf{PDF}s) for the neutrino signal, expressing different assumptions about the relative neutrino flux for each source. Each weighting scheme represents its own test of the data.

The analysis differs from previous point source searches most drastically in two points. \begin{enumerate}
 \item The blazar populations comprise nearly 2 orders of magnitude more sources.
 \item For the first time, we use a model-independent weighting scheme. In this test of the data, we make nearly no assumption about the exact $\nu/\gamma$ correlation, except that the neutrino flux originates from the defined blazar positions.
\end{enumerate}

Section \ref{section:blazar_populations} defines the five blazar populations considered in this analysis.
Section \ref{section:data} describes the muon track dataset used for this search. Section \ref{section:analysis} 
summarizes the analysis, including the technique of the unbinned stacking search, a description of different weighting schemes, 
the confidence interval construction, and a discussion on potential biases from non-hadronic contributions to the $\gamma$-ray flux. 
Section \ref{section:results} presents the analysis results and section \ref{section:discussion} discusses their implications.

\section{The blazar populations}
\label{section:blazar_populations}

The Fermi-LAT 2LAC catalogue \citep{Ackermann2011} contains 862 GeV-emitting blazars at high galactic latitudes $|b|>10^{\circ}$ that are not affected by potential source confusion\footnote{No source confusion means that the \texttt{CLEAN} flag from the
catalogue for a particular source is set.}. The data for this catalog was taken between August 2008 and August 2010. We use the spectroscopic classification into FSRQ and 
BL\,Lac objects \citep{Stickel1991} and the independent classification into LSP, ISP and HSP
objects \citep{Abdo2010} to define sub-populations of these 862 objects. 
We do not impose any other cuts (e.g. on the $\gamma$-ray flux) because the exact neutrino flux expectations are unknown as outlined in section \ref{sec:introduction}.
The motivations for the particular sub-samples are described in the following.

\begin{description}

\item[All 2LAC Blazars (862 objects)]{
The evolutionary blazar sequence \citep{Cavaliere2002, Bottcher2002} suggests that blazars form a continuous spectrum of objects that are connected via cosmological evolution. A recent study by \citet{Ajello2014} supports this hypothesis. Since the corresponding evolution of the neutrino emission is not known, the most unbiased assumption is to group all blazars together. 
This is especially justified for the 
analysis using the equal weighting scheme discussed in section \ref{section:weighting_schemes}.
}

\item[FSRQs (310 objects)]{
The class of FSRQs show strong, broad emission lines that potentially act as intense radiation targets for photomeson
production of neutrinos \citep{Atoyan2001,Murase2014}.
}

\item[LSPs (308 objects)]{
The majority of FSRQs are LSP objects. \citet{Giommi2012} argue that LSP-BL\,Lacs are actually physically similar to FSRQs, but whose emission lines are overwhelmed by the strong jet continuum. This sample therefore groups all LSP objects together.}

\item[ISP+HSPs (301 objects)]{
HSP objects differ from LSP objects in terms of their luminosities and mainly consist of BL\,Lacs \citep{Ajello2014}. 
The peak-frequency boundary between LSP and HSP is only defined artificially, with ISP objects filling the gap. 
In order to have a larger sample of objects, the HSP objects are grouped together
with the ISP objects in one combined sample for this analysis.}

\item[LSP-BL\,Lacs (68 objects)]{
Objects that share the LSP and BL\,Lac classification have been specifically considered for neutrino emission 
in \citet{Muecke2003}. Therefore we test them as a separate sample. They form the smallest sub population in this analysis. 
}
\end{description}

\begin{figure*}
\epsscale{1.1}
\plottwo{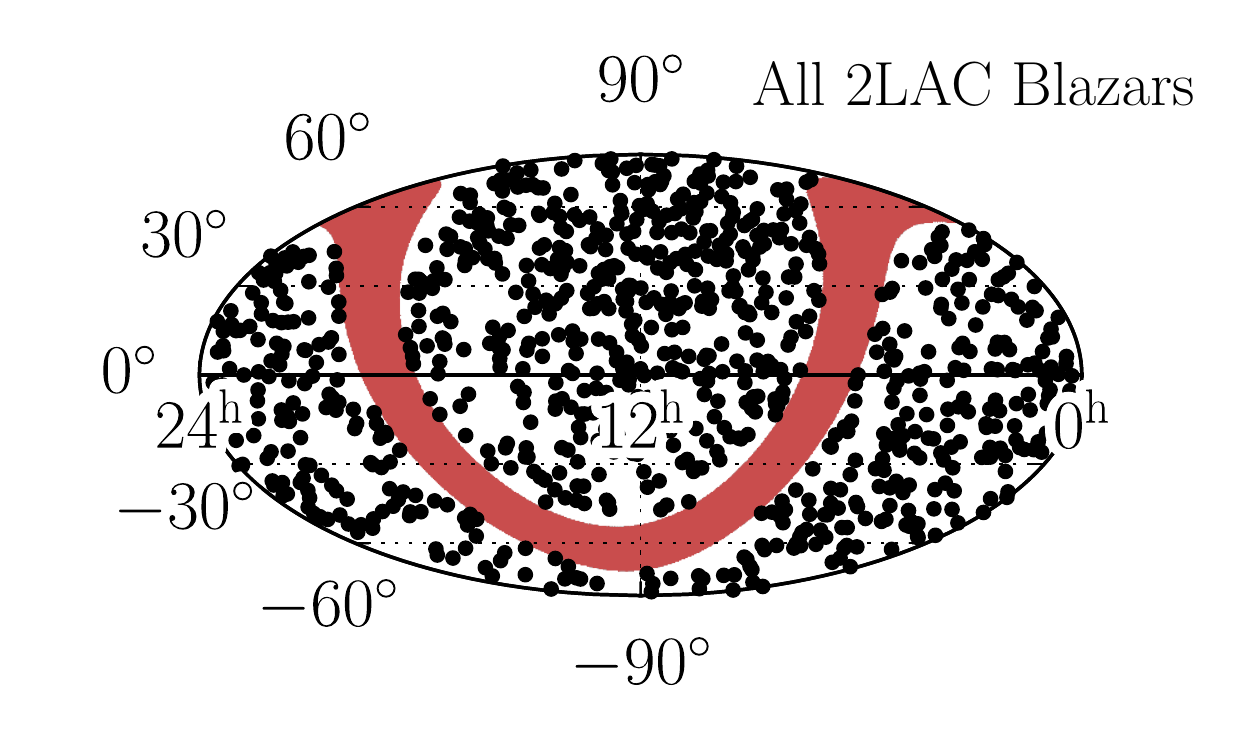}{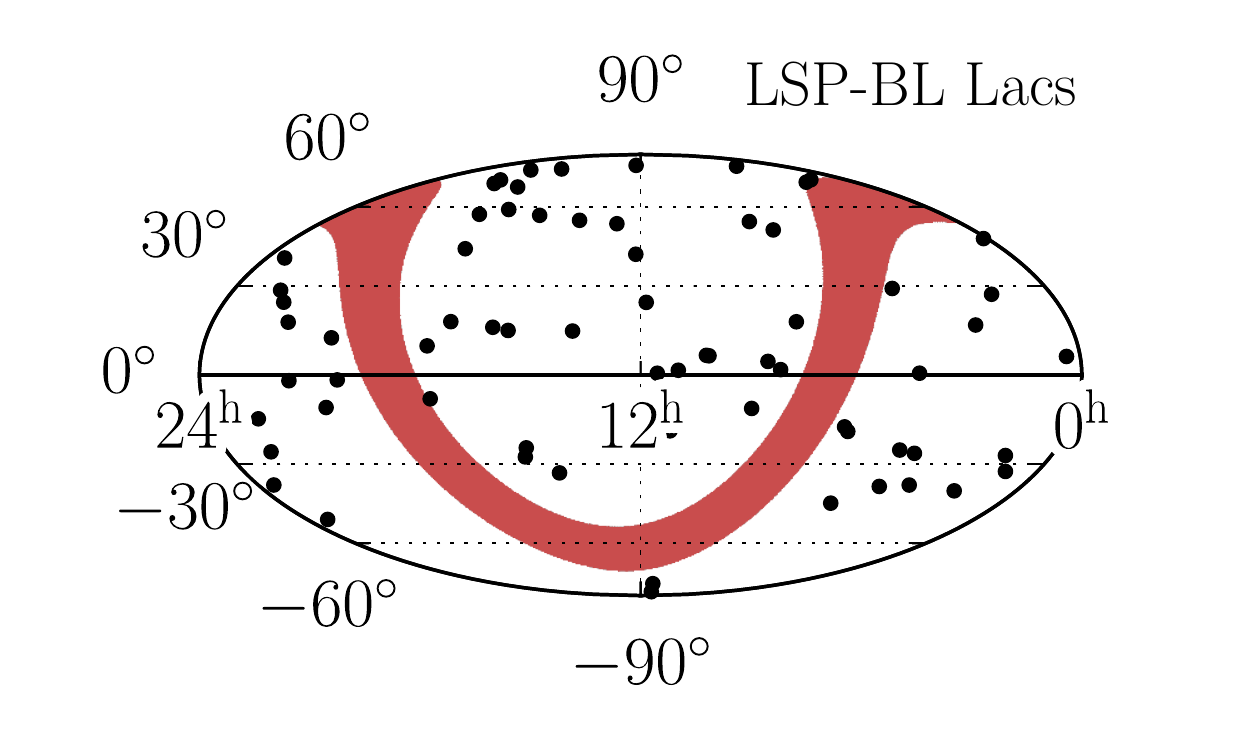}
\caption{Distribution of sources in the sky for the largest and smallest sample of blazars (in equatorial Mollweide projection) --- (left) The largest sample, all 2LAC blazars (862 sources) --- (right) The smallest sample, LSP-BL\,Lacs (68 sources). The excluded region of the catalogue ($|b|\leq 10 ^\circ$) is highlighted in red.}
\label{fig:blazar_distribution}
\end{figure*}

\begin{figure}
\epsscale{1.1}
\plotone{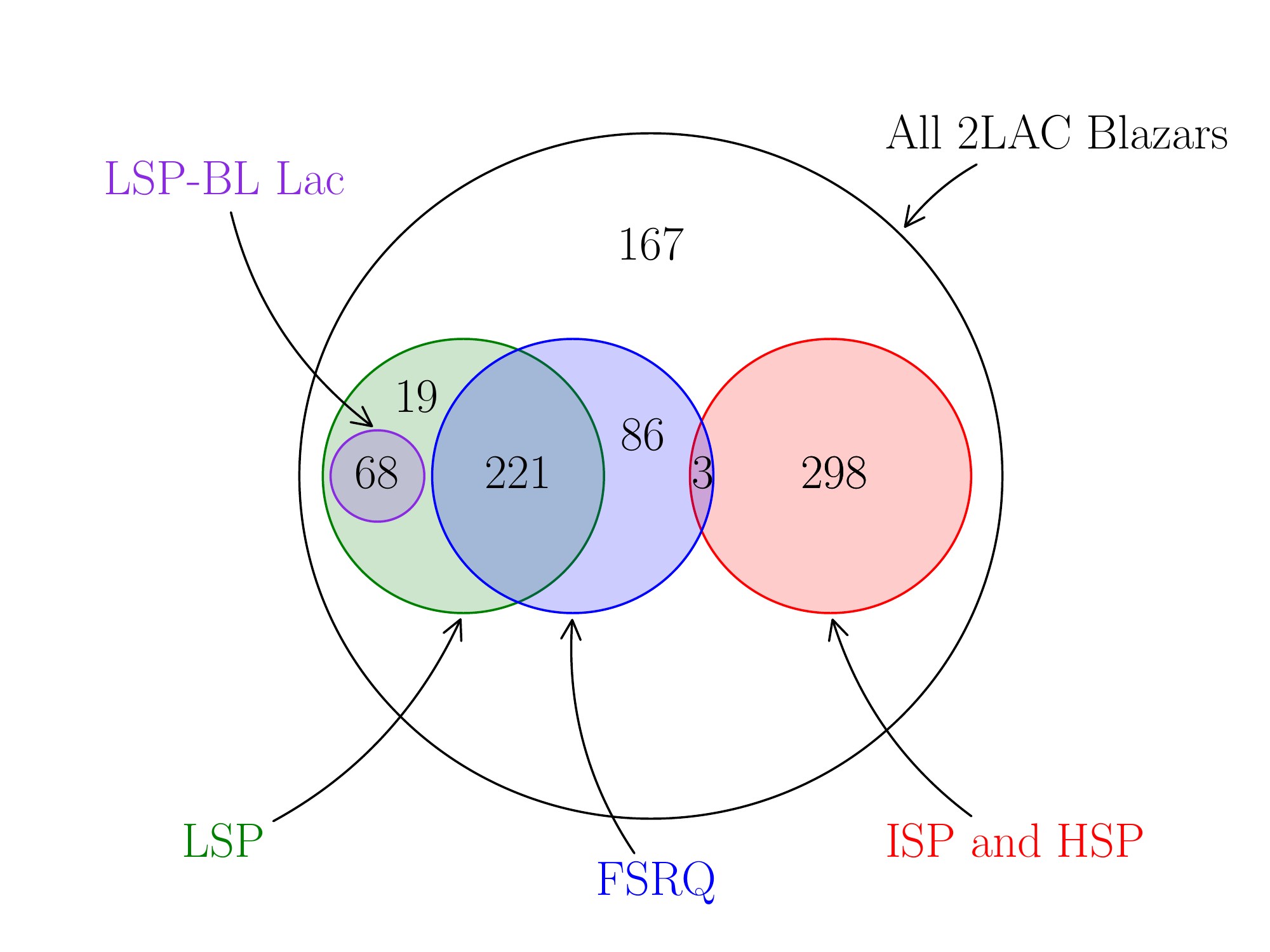}
\caption{Visualization of the source overlap between the different blazar populations.}
\label{fig:population_overlap}
\end{figure}

The distribution of the sources on the sky for the largest sample (all 2LAC blazars) and smallest sample (LSP-BL\,Lacs) are shown in 
figure \ref{fig:blazar_distribution}. A modest LAT-exposure deficit and lower sky-coverage by optical surveys in the southern sky lead to a slight deficit of objects in southern hemisphere \citep{Ackermann2011}. The effect is most prominent for the BL Lac-dominated samples. However, blazars without optical association are also included in the 2LAC catalog and partly make up for this asymmetry in the total sample. For simplicity, we assume a quasi-isotropic source distribution for all populations (excluding the source-free region around the galactic plane) for the calculation of quasi-diffuse fluxes. This assumption also seems reasonable looking at the weight distribution of sources (equal weighting) in figures \ref{fig:histogrammed_weights} (a)--(e), appendix \ref{appendix:supplementary}. Figure \ref{fig:population_overlap} shows the overlap between the samples. 
The LSP-BL\,Lac, FSRQ and ISP+HSP sample are nearly independent, with a small overlap of 3 sources between the FSRQs and ISP+HSP samples. The largest overlap exists between the FSRQ and LSP samples, which share around $60 \%$ of their sources. The all-blazar sample contains 167 sources that are not shared with any sub sample. These are sources that are either unclassified or only classified as BL\,Lac objects with no corresponding synchrotron peak classification.


\section{Data selection}
\label{section:data}
IceCube is a neutrino telescope located at the geographic South pole. It consists of about one $\mathrm{km}^3$ of Antarctic ice that is instrumented with 5160 optical photo sensors which are connected via cables (``strings'') with the data aqcuisition system at the surface. The photo sensors detect Cherenkov light emitted by charged particles that are produced in the interactions of neutrinos with nuclei and electrons in the glacial ice. The geometry and sensitivity of the photo sensors leads to an effective energy threshold for neutrinos of about $100 \ \mathrm{GeV}$. A more detailed description of the detector and the data acquisition can be found in \citet{Abbasi2009}.

Two main signatures can be distinguished for the recorded events: ``track-like'' and ``shower-like''. Only track-like events are of 
interest for the analysis here. They are the characteristic signature of muons produced in the charged-current interactions of muon neutrinos\footnote{We neglect track-like signals from $\nu_{\tau} + N \rightarrow \tau + X \rightarrow \mu + X$, i.e. muons as end products of a $\nu_{\tau}$ charged-current interaction chain. The $\tau \rightarrow \mu$ decay happens with a branching fraction of only $ 17 \%$ \citep{Olive2014}, and the additional decay step lowers the outgoing muon energy, leading to an even further suppression of the $\nu_\tau$ contribution in a sample of track-like events. For hard fluxes (spectral index 1-2) above PeV energies, where the $\nu_\tau$-influence becomes measurable due to $\nu_\tau$-regeneration \citep{Bugaev2004}, this treatment is conservative.}. 

IceCube was constructed between 2006 and 2011 with a final configuration of 86 strings. 
We use data from the 59-string (\textbf{IC-59}), 79-string \textbf{(IC-79}) and 86-string (\textbf{IC-86}) configurations of the IceCube detector recorded between May 2009 and April 2012. 
In contrast to previous publications, we do not include data from the 40-string configuration here since the ice model description in the IC-40 MonteCarlo datasets is substantially different and the sensitivity gain would be marginal. 
The track event selection for the three years of data is similar to the ones described in \citet{Aartsen2013a} and \citet{Aartsen2014}. The angular resolution of the majority of events in the track sample is better than $1^{\circ}$ for events with reconstructed muon energies above 10 TeV \citep{Aartsen2014}. The angular reconstruction uncertainty is calculated following the prescription given in \citet{Neunhoffer2006}. We apply one additional minor selection criterion for the estimated angular uncertainty of the reconstructed tracks ($\sigma_{est.} \leq 5^{\circ}$) for computational reasons. The removed events do not have any measurable effect on the sensitivity. 
Event numbers for the individual datasets are summarized in table \ref{table:dataset_statistics}. 

\begin{table}
\centering
\begin{tabular}{ c||c|c|c }
  \hline
  dataset & all sky & northern sky & southern sky  \\
  \hline
  \hline
  IC-59 &  107011 & 42781  & 64230  \\
  \hline
  IC-79 & 93720 &  48782  &  44938  \\
  \hline
  IC-86  & 136245   &  61325   & 74920   \\
  \hline
\end{tabular}
\caption{Total number of data events in the respective datasets of IC-59, IC-79 and IC-86 for each celestial hemisphere. ``Northern sky'' means the zenith angle $\theta$ for the incoming particle directions is equal to or larger than $90^{\circ}$. ``Southern sky'' means $\theta < 90^{\circ}$.  
}
\label{table:dataset_statistics}
\end{table}
The dataset is dominated by bundles of atmospheric muons produced in cosmic-ray air shower interactions for tracks coming from the southern hemisphere ($\theta < 90^{\circ}$). 
Tracks from the northern hemisphere ($\theta \geq 90^{\circ}$) originate mostly from atmospheric neutrino interactions that produce muons. 
In order to reduce the overwhelming background of direct atmospheric muons to an acceptable level, it is necessary to impose a high-energy cut for events from the southern hemisphere. The cut raises the effective neutrino energy threshold to approximately 100 TeV \citep{Aartsen2014}, reducing the sensitivity to neutrino sources in this region by at least 1 order of magnitude for spectra softer than $\mathrm{E}^{-2}$. Only for harder spectra, the southern sky has a significant contribution to the overall sensitivity.
The northern sky does not require such an energy cut, as upgoing tracks can only originate from neutrino interactions, which have a much lower incidence rate. However, at very high energies (again around $100 \mathrm{TeV}$), the Earth absorbs a substantial fraction of neutrinos, reducing also the expected astrophysical signal. Charged-current $\nu_\mu$-interactions can happen far outside the instrumented volume and still be detected, as high-energy muons may travel several kilometers through the glacial ice before entering the detector. This effect increases the effective detection area for certain arrival directions, mostly around the horizon. 

The most sensitive region is therefore around the celestial equator, which does not require a high energy cut, provides ample target material surrounding the detector, i.e. a large effective area, and does not suffer from absorption of neutrinos above $100$ TeV. 
 However, these zenith-dependent sensitivity changes are mostly important for the interpretation of the results (see e.g. section \ref{section:generic_upper_limits}). The likelihood approach takes these differences into account with the "acceptance" term in eq. (\ref{eq:weighting_term}), section \ref{section:llh}, and a separation into several zenith-dependent analyses is not necessary. For more details on the properties of the datasets and the zenith-dependent sensitivity behaviour, we refer to \citet{Aartsen2013a} and \citet{Aartsen2014}.

\section{Analysis}
\label{section:analysis}
\subsection{The likelihood function for unbinned ML stacking}
\label{section:llh}

The analysis is performed via an extended unbinned maximum likelihood fit \citep{Barlow1990}. The likelihood function consists 
of two PDFs, one PDF $B(\overline{x})$ for a background hypothesis and one PDF $S(\overline{x})$ for a signal hypothesis. 
Requiring the total number of observed events to be the sum of the signal and background events, the log-likelihood function can be written as
\begin{equation}
\begin{split}
\mathrm{ln}(L)\{n_s, \Gamma_{\mathrm{SI}}\} &= \sum_{i=1}^{N} \mathrm{ln} 
\left(  \frac{n_s}{N} \cdot S(\delta_i, RA_i, \sigma_i, \varepsilon_i; \Gamma_{\mathrm{SI}}) 
\right. \\ &\left. + \left(1-\frac{n_s}{N} \right)\cdot B(\mathrm{sin}(\delta_i), \varepsilon_i)  \right) ,
\end{split}
\label{eq:llh}
\end{equation}
where $i$ indexes individual neutrino events.
The likelihood function depends on two free parameters: the normalization factor $n_s$ and spectral index $\Gamma_{\mathrm{SI}}$ of the total blazar signal. For computational reasons we assume that each source of a given population shares the same spectral index. 
The background evaluation for each event depends on the reconstructed declination $\delta_i$
and the reconstructed muon energy $\varepsilon_i$. The signal part additionally depends on the reconstructed right ascension $\mathrm{RA}_i$, the angular error estimator $\sigma_i$ and the power-law spectral index $\Gamma_{\mathrm{SI}}$. 

The background PDF is constructed from binning the recorded data in reconstructed declination and energy. It is evaluated as 
\begin{equation}
B(\mathrm{sin}(\delta_i), \varepsilon_i)=\frac{1}{2 \pi} \cdot f(\mathrm{sin}(\delta_i), \varepsilon_i),
\end{equation}
where $\frac{1}{2 \pi}$ arises from integration over the right ascension and $f$ is the normalized joint probability distribution of the events in declination $\mathrm{sin}(\delta)$ and energy $\varepsilon$.

The signal PDF that describes a given blazar population is a superposition of the individual PDFs for each source,
\begin{equation}
\begin{split}
&S(\delta_i, \mathrm{RA}_i, \sigma_i, \varepsilon_i; \Gamma_{\mathrm{SI}}) 
\\ &= \frac{\sum_{j=1}^{N_{src}}{w_{j} \cdot S_j(\delta_i, \mathrm{RA}_i, \sigma_i, \varepsilon_i; \Gamma_{\mathrm{SI}})}}{\sum_{j=1}^{N_{src}}{w_{j}}} ,
\end{split}
\label{eq:cumulative_signal}
\end{equation}
where $w_j$ is a weight determining the relative normalization of the PDF $S_j$ for source $j$. This weight therefore accounts for the relative contribution of source $j$ to the combined signal. In general, different choices of $w_j$ are possible. 
The two choices used in this work are discussed in section \ref{section:weighting_schemes}. Each term $S_j$ in equation \ref{eq:cumulative_signal} is evaluated as
\begin{equation}
\begin{split}
 &S_j(\delta_i, \mathrm{RA}_i, \sigma_i, \varepsilon_i; \Gamma_{\mathrm{SI}}) \\ =  \frac{1}{2\pi {\sigma_i}^2}\cdot &\mathrm{exp}\left( \frac{1}{2} \cdot \left(\frac{ \Psi_{ij}[\delta_i, \mathrm{RA}_i]}{\sigma_i}\right)^2\right) \cdot g_j(\varepsilon_i; \Gamma_{\mathrm{SI}}) ,
\end{split}
\end{equation}
where the spatial term is expressed as a 2D symmetric normal distribution and $g_j$ is the normalized PDF for the reconstructed muon
energy for source $j$. The term $\Psi_{ij}$ is the angular
separation between event $i$ and source $j$. 

\subsection{Weighting Schemes}
\label{section:weighting_schemes}

The term $w_j$ in equation \ref{eq:cumulative_signal} parametrizes the relative contribution of source $j$ to the combined signal. It corresponds to the expected number of events for source $j$, which can be expressed as
\begin{equation}
w_{j} =  \int_{E_{\nu,min}}^{E_{\nu,max}}{  \Phi_{0,j} \cdot h_j(E_{\nu}) \cdot A_{\mathrm{eff}}(\theta_{j} , E_{\nu} ) \ dE_{\nu}  } ,
\end{equation}
where $A_{\mathrm{eff}}(\theta_{j}, E_{\nu})$ is the effective area for incoming muon neutrinos from a given source direction at a given energy,
$h_j(E_{\nu})$ denotes the normalized neutrino energy spectrum for source $j$, and $\Phi_{0,j}$ its overall flux normalization. The integration bounds $E_{\nu,\mathrm{min}}$ and $E_{\nu,\mathrm{max}}$ are set to $10^{2} \ \mathrm{GeV}$ and $10^9 \ \mathrm{GeV}$ respectively, except for the differential analysis (see  section \ref{section:statistical_tests}), in which they are defined for the given energy band. 

Under the assumption that all sources share the same spectral power-law shape, 
$w_j$ further simplifies via
\begin{align}
w_{j}  &=  \left[\Phi_{0,j}\right] \cdot \left[\int_{E_{\nu,min}}^{E_{\nu,max}}{   h(E_\nu; \Gamma_{\mathrm{SI}}) \cdot A_{\mathrm{eff}}(\theta_{j} , E_{\nu} ) \ dE_{\nu}  } \right] \nonumber \\  &= \left[C \cdot w_{j, \mathrm{model}} \right]  \cdot \left[w_{j, acc.}(\theta_{j}, \Gamma_{\mathrm{SI}})\right] ,
\label{eq:weighting_term}
\end{align}
and splits into a ``model'' term, where $w_{j, \mathrm{model}}$ is proportional to the expected relative neutrino flux of source $j$, and into an ``acceptance'' term, which is fixed by the position of the source and the global energy spectrum. 
The term $w_{j, \mathrm{model}}$ is not known, and its choice defines the ``weighting scheme'' for the stacking analysis.
The following two separate weighting schemes are used for the signal PDF in the likelihood analysis, leading to two different sets of tests.

\subsubsection{$\gamma$-weighting}
\label{sec:gamma_weighting}

For this weighting scheme we first have to assume that the $\gamma$-ray flux can be modeled as being quasi-steady between 2008 and 2010, the time period which forms the basis for the 2LAC catalog. This makes it possible to extrapolate the flux expectation of each source to other time periods, e.g. into the non-overlapping part of the data-taking period of the IceCube data for this analysis (2009-2012).
Each model weight, i.e. the relative neutrino flux expected to arrive from a given source, is then given by the source's $\gamma$-ray energy flux observed by Fermi-LAT in the energy range between $E> 100 \ \mathrm{MeV}$ and $E> 100 \ \mathrm{GeV}$.
\begin{equation}
w_{j, \mathrm{model}}=\int_{100 \mathrm{MeV}}^{100 \mathrm{GeV}}{E_{\gamma} \frac{d \phi_{\gamma, j}}{d E_{\gamma}} \ dE_{\gamma}}
\end{equation} 
This is motivated by the fact that a similar amount of energy is channeled into the neutrino and $\gamma$-ray emission if pion decay from $pp$ or $p\gamma$ interactions dominates the high-energy interaction. While the source environment is transparent to high-energy neutrinos, it might not be for $\gamma$-rays. Reprocessing of $\gamma$-rays due to $\gamma \gamma$ interactions might then shift the energies of the photons to GeV and sub-GeV energies before they can leave the sources, which would make them detectable by the Fermi-LAT. This might even be expected in $p\gamma$ scenarios \citep{Murase2015}. Since a large fraction of blazars are located at high redshifts $z\geq1$ \footnote{With the exception of HSP objects, see \citet{Ackermann2011}.} , this reprocessing will also take place during propagation of the photons in the extragalactic background light (\textbf{EBL}),
shifting $\gamma$-ray energies below a few hundred GeV for such sources \citep{Dominguez2013}. This places them potentially again in the energy range of the Fermi-LAT 2LAC catalogue. 
Even in the case that synchrotron contributions (e.g. muon or pion synchrotron radiation) dominate over pion decay in the $\mathrm{MeV}$-$\mathrm{GeV}$ range, 
which has been considered in particular for BL\,Lac objects \citep{Muecke2003}, one would expect the overall $\gamma$-ray emission to be proportional to the neutrino emission. This is also the case in models where inverse Compton processes dominate the high-energy $\gamma$-ray emission \citep{Murase2014}. 

The preceding arguments in favour of a $\gamma$-weighting scheme assume that all sources show equal proportionality. On a source-by-source basis, however, the proportionality factor can vary, as already mentioned in section \ref{sec:introduction}.

One contributing factor is the fact that Fermi probes different sections of the blazar $\gamma$-ray peak for each source relative to the peak position. For simplicity, we do not perform a spectral source-by-source fit in this paper, leaving this aspect for potential future work. This is also mostly an issue for the "All 2LAC-Blazar" sample, since the other sub-classifications described in section \ref{section:blazar_populations} depend on the peak position and this effect is largely mitigated.
There are additional reasons for source-by-source fluctuations in the $\gamma/\nu$ correlation due to EBL reprocessing. First, the EBL absorption might not be sufficient for close-by sources, such that emerging high-energy $\gamma$-rays are not reprocessed into the energy range of the 2LAC catalogue which ends at $100 \ \mathrm{GeV}$.
Second, EBL reprocessing differs between sources depending on the line-of-sight magnetic fields which deflect charged particle pairs produced in EBL cascades \citep{Aharonian1994} differently. 
Third, strong in-source $\gamma \gamma$ reprocessing could lead to $\gamma$-rays at even lower energies than $100 \ \mathrm{MeV}$ \citep{Murase2015} which would be below the 2LAC energy range. 

All results presented in section \ref{section:results} making use of the $\gamma$-weighting scheme assume that the potential source-to-source fluctuations in the $\gamma-\nu$ correlation described here average out for large source populations and can be neglected. More information on the distribution of the weights in dependence of declination can be found in figures \ref{fig:histogrammed_weights} (a)--(e), appendix \ref{appendix:supplementary}.



\subsubsection{Equal weighting}


The $\gamma$-weighting scheme is optimal under the assumption that the neutrino flux follows the measured $\gamma$-energy flux exactly. Given the the uncertainties discussed in section \ref{sec:gamma_weighting}, we also use another weighting scheme,
\begin{equation}
w_{j,\mathrm{model}}=1 ,
\end{equation}
which we expect to be more sensitive eventually if the actual $\gamma-\nu$ correlation varies strongly from source to source. It provides a complementary and model-independent test in which we are maximally agnostic to the degree of correlation between $\gamma$-ray and neutrino luminosities. 

We do not assume a specific neutrino emission in a given source when calculating the flux upper limits for the equal weighting scheme, in particular no equal emission. We only assume, to some approximation, that the differential source count distributions (\textbf{SCD}s) of $\gamma$-rays and neutrinos have comparable shapes. The differential source count distribution, $dN/dS$, describes how the energy-integrated flux $S$ is distributed over all sources and is a crucial property of any cosmological source population. 
Section \ref{section:ensemble_simulations} provides more information on the technical aspects of neutrino flux injection in the equal weighting test. Appendix \ref{appendix:scd_dependence} then discusses why the methodology is robust against variations in the actual shape of the $dN/dS$ distribution for the neutrino flux in the IceCube energy range, and why the final result is valid even if the neutrino SCD is different from the $\gamma$-ray SCD.

\subsection{Statistical tests}
\label{section:statistical_tests}

We perform statistical tests for each population of blazars.
The log-likelihood difference $\lambda$ defines our test statistic (\textbf{TS}), given by
\begin{equation}
\begin{split}
\lambda= &-2 \cdot \mathrm{log}(L)\{n_s=0\} \\ &+ 2 \cdot \mathrm{log}(L)\{n_s=n_{s,\mathrm{max}}, \Gamma_{\mathrm{SI}}=\Gamma_{\mathrm{SI},\mathrm{max}}\} ,
\end{split}
\end{equation}
where $n_{s,\mathrm{max}}$ and $\Gamma_{\mathrm{SI},\mathrm{max}}$ 
are the number of signal events and the signal spectral index that maximize 
the TS. We simulate an ensemble of background-only skymaps, where the TS distribution is compared with the TS value obtained from the data. The p-value is then defined as the fraction of skymaps in the background ensemble that has a larger TS value than the one observed. Ensembles of skymaps with different injected signal strengths are then used to calculate the resulting confidence interval. 
See section \ref{section:ensemble_simulations} for details on the skymap simulations. 

In total we perform two distinct types of tests for which p-values are calculated. The first (``integral'') assumes a power law spectrum for the blazar emission over the full energy range observable with IceCube (unless stated otherwise). 
The second (``differential'') assumes a neutrino signal that is confined to a small energy range (half a decade in energy) and has a power law spectrum with a spectral index of $-2$ within this range. We perform the differential test for 14 energy ranges between $100 \ \mathrm{GeV}$ and $1 \ \mathrm{EeV}$. 

\subsection{Simulations}
\label{section:ensemble_simulations}

We estimate the sensitivity of our searches in both weighting schemes using an ensemble of simulated skymaps containing both background and signal events. 

We simulate the background by drawing events from the experimental data sample, then randomizing their right ascensions to remove any correlation with blazar positions. This is the same method used in previous IceCube point source searches \citep{Aartsen2013a,Aartsen2014} and mitigates systematic uncertainties in the background description due to the data-driven event injection. 

The injection for signal differs depending on the weighting schemes. For the $\gamma$-weighting scheme, we inject signal events with the relative flux contribution of each source determined by the weight factors $w_{j,\mathrm{model}}$ that are used in the PDF.
In the equal weighting scheme, following the same approach would lead to a simulated signal of $n$ equally bright sources, which 
is not realistic for a population distributed widely in redshift and luminosity. Therefore
we inject events using a relative neutrino flux contribution that follows a realistic SCD. Since the neutrino $dN/dS$ distribution of blazars is unknown, we have chosen to use the blazar $\gamma$-ray SCD published in \citet{Abdo2010a} as a template\footnote{This blazar SCD strictly stems from the 1FGL catalogue \citep{Abdo2010b}, but any SCD based on a newer catalog is not expected to change significantly since a large fraction of the total $\gamma$-ray flux is already resolved in the 1FGL.}.
Here we assume that for the population under investigation, the relative contributions to the total neutrino flux
are distributed in a similar fashion as the relative contributions to the total $\gamma$-ray flux. 
However, there are no assumptions about the correlation of the neutrino and $\gamma$-ray flux for individual sources. 

There are two reasons to choose the $\gamma$-ray SCD as the primary template for the shape of the neutrino SCD. The first is that we select the populations based on their $\gamma$-ray emission to start with. 
The second is that the form of the high-energy $\gamma$-ray SCD is quite general, and has also been observed with AGN detected in the radio \citep{Hopkins2003}  and x-ray \citep{Georgakakis2008a} bands. It starts with quasi-Euclidean behavior ($S^{5/2} \cdot dN/dS \approx \mathrm{const.}$) at high fluxes, 
and then changes to a harder power law index towards smaller flux values which ensures that the total flux from the population remains finite.

The skymap simulations are performed for many possible SCD realizations by sampling from the $dN/dS$ distribution. This is necessary since the number of signal events expected in IceCube for a given neutrino flux varies greatly over the two hemispheres (see section \ref{section:data}). Thus, it matters how the neutrino flux is distributed over the individual sources for the value of the resulting confidence interval. 
The shape of the SCD and the flux sampling range have an additional impact. See appendix  \ref{appendix:scd_dependence} for further details in the context of confidence interval construction.



\section{Results}
\label{section:results}


\subsection{Observed p-values}

Table \ref{table:integral_pvalues} summarizes p-values 
for the ``integral'' test (see section \ref{section:statistical_tests}). Nine out of the ten tests show over-fluctuations, but no significant excess. 
We find the strongest over-fluctuation, a $6 \%$ p-value, using the equal-weighting scheme for all 2LAC blazars. We omit a trial-factor correction because the populations have a large overlap and the result is not significant. 

\begin{figure}
\epsscale{1.1}
\plotone{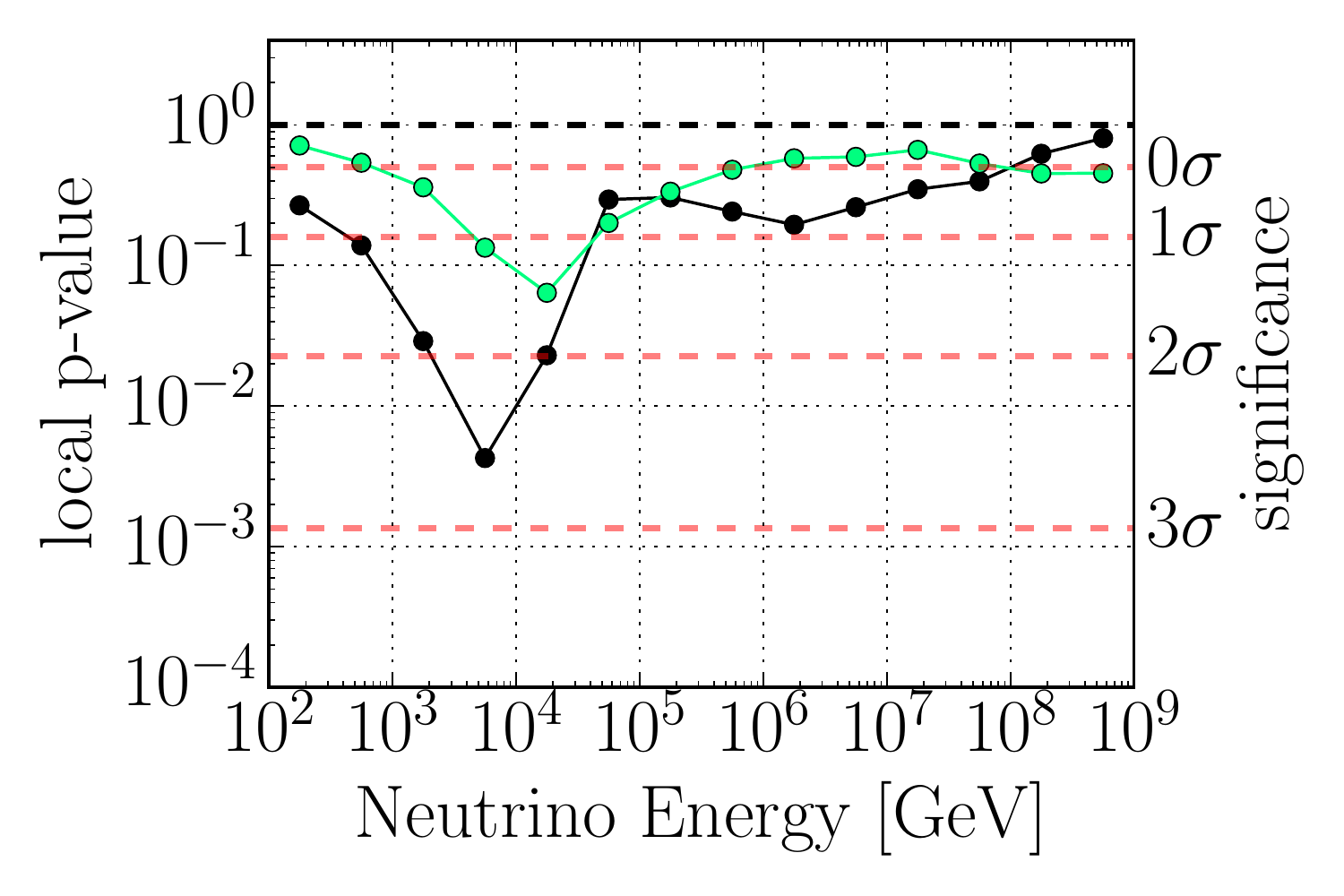}
\caption{Local p-values for the sample containing all 2LAC blazars using the equal-weighting scheme (black) and  $\gamma$-weighting scheme (green) in the differential test.}
\label{fig:differential_pvalues}
\end{figure}

Figure \ref{fig:differential_pvalues} shows the p-values from the corresponding ``differential'' test. The largest excess is visible in the $5-10 \ \mathrm{TeV}$ energy band with a pre-trial p-value of $4 \cdot 10^{-3}$. This outcome is totally compatible with a fluctuation of the background, since the effect of multiple trials has to be taken into account which reduces the significance of the observation substantially. An accurate calculation of the trial-corrected p-value is again difficult, as neither the five blazar samples, nor the 14 tested energy ranges per sample, are independent. We again omit it for simplicity. 

Comparing the differential p-value plot of all 2LAC blazars with the other populations (see figures \ref{fig:differential_limits_and_pvalues} (a)--(e) in appendix \ref{appendix:supplementary}), one finds that the overfluctuation is caused by the LSP-BL Lac-, FSRQ- and ISP/HSP population, which are nearly independent and each show a small excess in 5 TeV - 20 TeV region. In the $\gamma$-weighting scheme, the ISP/HSP p-value distribution is nearly flat, which leads to the weaker overfluctuation in the "all 2LAC blazar" sample compared to the equal weighting scenario.

\begin{table}
\centering
\begin{tabular}{ c||c|c }
\hline
\multirow{2}[0]{*}{Population} & \multicolumn{2}{c}{p-value} \\
& $\gamma$-weighting & equal weighting \\
 \hline
 \hline
All 2LAC blazars & $36 \%$ $(+0.4 \sigma)$ & $6 \%$ $(+1.6 \sigma)$  \\
\hline
FSRQs & $34 \%$ $(+0.4 \sigma)$ & $34 \%$ $(+0.4 \sigma)$  \\
\hline
LSPs & $36 \%$ $(+0.4 \sigma)$ & $28 \%$ $(+0.6 \sigma)$ \\
\hline
ISP/HSPs & $>50 \%$ & $11 \%$ $(+1.2 \sigma)$  \\
\hline
LSP-BL\,Lacs & $13 \%$ $(+1.1 \sigma)$ & $7 \%$ $(+1.5 \sigma)$ \\
\hline
 \end{tabular}
\caption[P-values for the spectrum integrated search]{P-values and the corresponding significance in units of standard normal deviations in the power-law test. The table shows the results for both weighting schemes.
The values do not include a trial factor correction.}
\label{table:integral_pvalues}
\end{table}

\subsection{Flux upper limits}
\label{section:upper_limits}

Since no statistically significant neutrino emission from the analyzed source populations was found, we calculate flux upper limits using various assumptions about their energy spectrum. We use the $CL_s$ upper limit construction \citep{Read2000}. It is more conservative than a standard Neyman construction, e.g. used in \citet{Aartsen2014}, but allows for a proper evaluation of under-fluctuations of the background which is used for the construction of differential flux upper limits.

We give all further results in intensity units and calculate the quasi-diffuse flux\footnote{The flux divided by the solid angle of the sky above 10 degrees galactic latitude, i.e. $0.83 \times 4 \pi$. See section \ref{section:blazar_populations} for a justification.} for each population. 

The flux upper limits in the equal weighting scheme are calculated using multiple samplings from an assumed neutrino SCD for the blazars, as already outlined in section \ref{section:ensemble_simulations}. Please refer to appendix \ref{appendix:scd_dependence} 
for further details about the dependence of the flux upper limit on the choice of the SCD and a discussion of the robustness of the equal-weighting results. In general, the equal weighting upper limit results do not correspond to a single flux value, but span a range of flux values. 

For each upper limit\footnote{With the exception of the differential upper limit.} we determine the valid energy range according to the procedure in appendix \ref{appendix:energy_range}. This energy range specifies where IceCube has exclusion power for a particular model, and is also used for visualization purposes in all figures.

Systematic effects influencing the upper limits are dominated by uncertainties on the absorption and scattering properties of the Antarctic ice and the detection efficiency of the optical modules.
Following \citet{Aartsen2014}, the total systematic uncertainty on the upper limits is estimated to be $21 \%$. Since we are dealing with upper limits only, we conservatively include the uncertainty additively in all figures and tables.

\subsection{Generic upper limits}
\label{section:generic_upper_limits}

Table \ref{table:generic_upper_limits} shows flux upper limits assuming a generic power-law spectrum for the tested blazar populations, calculated for the three different spectral indices $-1.5$, $-2.0$, and $-2.7$.

\begin{table}
\begin{tabular}{ c|c|c }

\hline
\multicolumn{3}{c}{ \rule{0pt}{2.5ex} Spectrum:  $\Phi_0 \cdot (E/\mathrm{GeV})^{-1.5}$   }\\
 
                \hline
          \multirow{2}[3]{*}{Blazar Class} & 
          \multicolumn{2}{c}{\rule{0pt}{3ex} $  {\Phi_{0}}^{90 \%} [\mathrm{GeV}^{-1} \mathrm{cm}^{-2} \mathrm{s}^{-1} \mathrm{sr}^{-1}]$} \\
            &  $\gamma$-weighting &  equal weighting \\
          \hline
          \hline
        \rule{0pt}{2.5ex} All 2LAC Blazars & {\color{black}$1.6 \times 10^{-12}$} & {\color{black}$4.6 \ (3.8 - 5.3) \times 10^{-12}$} \\ 
      \cline{1-3}
       \rule{0pt}{2.5ex} FSRQs &  {\color{black}$0.8 \times 10^{-12}$} & {\color{black}$2.1 \ (1.0 - 3.1) \times 10^{-12}$} \\
      \cline{1-3}
       \rule{0pt}{2.5ex} LSPs &  {\color{black}$1.0 \times 10^{-12}$} & {\color{black}$1.9 \ (1.2 - 2.6) \times 10^{-12}$} \\
      \cline{1-3}
       \rule{0pt}{2.5ex} ISPs/HSPs &  {\color{black}$1.8 \times 10^{-12}$} & {\color{black}$2.6 \ (2.0 - 3.2) \times 10^{-12}$} \\
      \cline{1-3}
       \rule{0pt}{2.5ex} LSP-BL\,Lacs &  {\color{black}$1.1 \times 10^{-12}$} & {\color{black}$1.4 \ (0.5 - 2.3) \times 10^{-12}$} \\
      \cline{1-3}
      
              \hline\hline
\multicolumn{3}{c}{ \rule{0pt}{2.5ex} Spectrum:  $\Phi_0 \cdot (E/\mathrm{GeV})^{-2.0}$   }\\
 
                \hline
          \multirow{2}[3]{*}{Blazar Class} & 
          \multicolumn{2}{c}{\rule{0pt}{3ex} $  {\Phi_{0}}^{90 \%} [\mathrm{GeV}^{-1} \mathrm{cm}^{-2} \mathrm{s}^{-1} \mathrm{sr}^{-1}]$} \\
            &  $\gamma$-weighting &  equal weighting \\
          \hline
          \hline
        \rule{0pt}{2.5ex} All 2LAC Blazars & {\color{black}$1.5 \times 10^{-9}$} & {\color{black}$4.7 \ (3.9 - 5.4) \times 10^{-9}$} \\ 
      \cline{1-3}
       \rule{0pt}{2.5ex} FSRQs &  {\color{black}$0.9 \times 10^{-9}$} & {\color{black}$1.7 \ (0.8 - 2.6) \times 10^{-9}$} \\
      \cline{1-3}
       \rule{0pt}{2.5ex} LSPs &  {\color{black}$0.9 \times 10^{-9}$} & {\color{black}$2.2 \ (1.4 - 3.0) \times 10^{-9}$} \\
      \cline{1-3}
       \rule{0pt}{2.5ex} ISPs/HSPs &  {\color{black}$1.3 \times 10^{-9}$} & {\color{black}$2.5 \ (1.9 - 3.1) \times 10^{-9}$} \\
      \cline{1-3}
       \rule{0pt}{2.5ex} LSP-BL\,Lacs &  {\color{black}$1.2 \times 10^{-9}$} & {\color{black}$1.5 \ (0.5 - 2.4) \times 10^{-9}$} \\
      \cline{1-3}
      
              \hline\hline
\multicolumn{3}{c}{ \rule{0pt}{2.5ex} Spectrum:  $\Phi_0 \cdot (E/\mathrm{GeV})^{-2.7}$   }\\
 
                \hline
          \multirow{2}[3]{*}{Blazar Class} & 
          \multicolumn{2}{c}{\rule{0pt}{3ex} $  {\Phi_{0}}^{90 \%} [\mathrm{GeV}^{-1} \mathrm{cm}^{-2} \mathrm{s}^{-1} \mathrm{sr}^{-1}]$} \\
            &  $\gamma$-weighting &  equal weighting \\
          \hline
          \hline
        \rule{0pt}{2.5ex} All 2LAC Blazars & {\color{black}$2.5 \times 10^{-6}$} & {\color{black}$8.3 \ (7.0 - 9.7) \times 10^{-6}$} \\ 
      \cline{1-3}
       \rule{0pt}{2.5ex} FSRQs &  {\color{black}$1.7 \times 10^{-6}$} & {\color{black}$3.3 \ (1.6 - 5.1) \times 10^{-6}$} \\
      \cline{1-3}
       \rule{0pt}{2.5ex} LSPs &  {\color{black}$1.6 \times 10^{-6}$} & {\color{black}$3.8 \ (2.4 - 5.2) \times 10^{-6}$} \\
      \cline{1-3}
       \rule{0pt}{2.5ex} ISPs/HSPs &  {\color{black}$1.6 \times 10^{-6}$} & {\color{black}$4.6 \ (3.5 - 5.6) \times 10^{-6}$} \\
      \cline{1-3}
       \rule{0pt}{2.5ex} LSP-BL\,Lacs &  {\color{black}$2.2 \times 10^{-6}$} & {\color{black}$2.8 \ (1.0 - 4.6) \times 10^{-6}$} \\
      \cline{1-3}
      
              \hline\end{tabular}
\caption{$90 \% $ C.L. upper limits on the diffuse ($\nu_\mu+\overline{\nu}_\mu$)-flux from the different blazar populations tested. The table contains results for power-law spectra with spectral indices $-1.5$, $-2.0$, and $-2.7$. The equal-weighting column shows the median flux upper limit
and the $90 \%$ central interval of different sample realizations of the Fermi-LAT source count contribution (in parentheses). All values include systematic uncertainties.}
\label{table:generic_upper_limits}
\end{table}
The distribution of the $\gamma$-ray energy flux among the sources in each population governs the flux upper limit in the $\gamma$-weighting scheme.  
It is mostly driven by the declination of the strongest sources in the population, due to the strong declination dependence of IceCube's effective area. For FSRQs, the two sources with the 
largest $\gamma$-weights (\texttt{3C 454.3} at $\mathrm{DEC}_{2000}=16^{\circ}$ and \texttt{PKS1510-08} at $\mathrm{DEC}_{2000}=-9^{\circ}$) 
carry around $15 \%$ of the total $\gamma$-weight of all FSRQs. Their positions close to the equator place them in the most sensitive region 
for the IceCube detector, and the $\gamma$-weighting upper limits for FSRQs are more than a factor of 2 lower than the corresponding equal-weighting limits. For the LSP-BL\,Lacs, the two strongest sources (\texttt{PKS 0426-380} at $\mathrm{DEC}_{2000}=-38^{\circ}$ 
and \texttt{PKS 0537-441} at $\mathrm{DEC}_{2000}=-44^{\circ}$) carry nearly $30 \%$ of the total $\gamma$-weight but are located in the southern sky, where IceCube is not very sensitive. The $\gamma$-weighting upper limit is therefore comparable to the equal-weighting upper limit. The reader is referred to appendix \ref{appendix:supplementary} for more information on the weight distribution. 

\begin{figure}
\epsscale{1.1}
\plotone{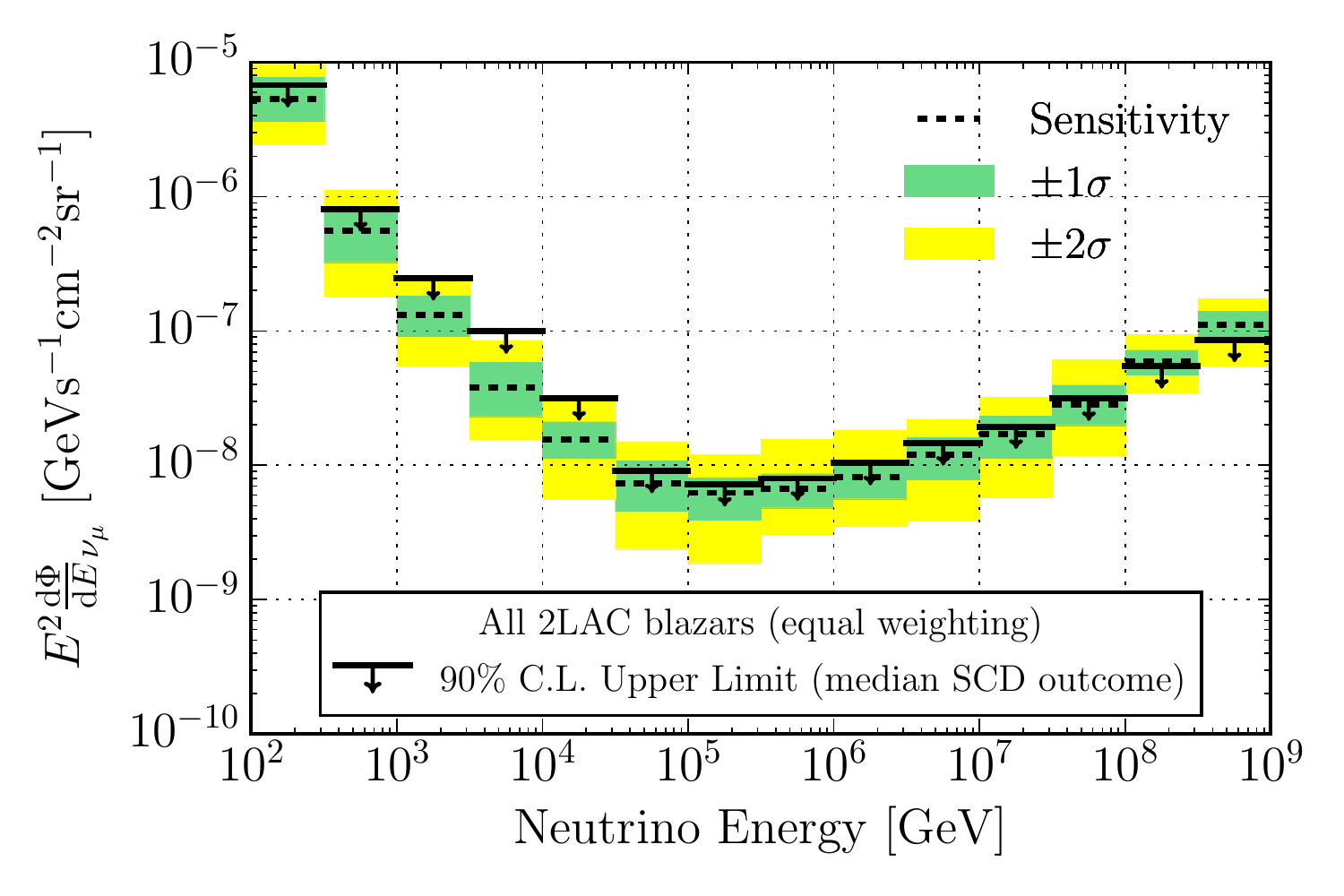}
\caption{Differential $90 \% \ \mathrm{C.L.}$ upper limit on the ($\nu_\mu + \overline{\nu}_\mu$)-flux using equal weighting for all 2LAC blazars. 
The $\pm 1 \sigma$ and $\pm 2 \sigma$ null expectation is shown in green and yellow, respectively. The upper limit and expected regions correspond to the median SCD sampling outcome.}
\label{fig:diff_upper_limit_brazilian}
\end{figure}

Figure \ref{fig:diff_upper_limit_brazilian} shows the differential upper limit in comparison to the median sensitivity for all 2LAC blazars using the equal-weighting scheme. This population showed the largest overfluctuation. We plot here the upper limit derived from the median SCD sampling outcome, since in general the equal weighting upper limit depends on the neutrino flux realization of the SCD (see appendix \ref{appendix:scd_dependence}). As expected, the differential limit is slightly higher, by a factor of about 2, than the median outcome in the energy range between 5~TeV and 10~TeV where the largest excess is observed. This is the average behavior for a soft flux with spectral index of about $-3.0$ \footnote{This can be read off in figure \ref{fig:energy_interval}. The ratio function indicates in which energy range a given flux function appears first, on average.}, if one assumes a simple power-law fit to explain the data. While such a physical interpretation can not be made yet, it will be interesting to observe this excess with future IceCube data. For information on the differential upper limits from the other samples the reader is referred to appendix \ref{appendix:supplementary}.

\subsection{The maximal contribution to the diffuse astrophysical flux}
\label{section:contribution_to_diffuse}

\begin{figure}
\epsscale{1.1}
\plotone{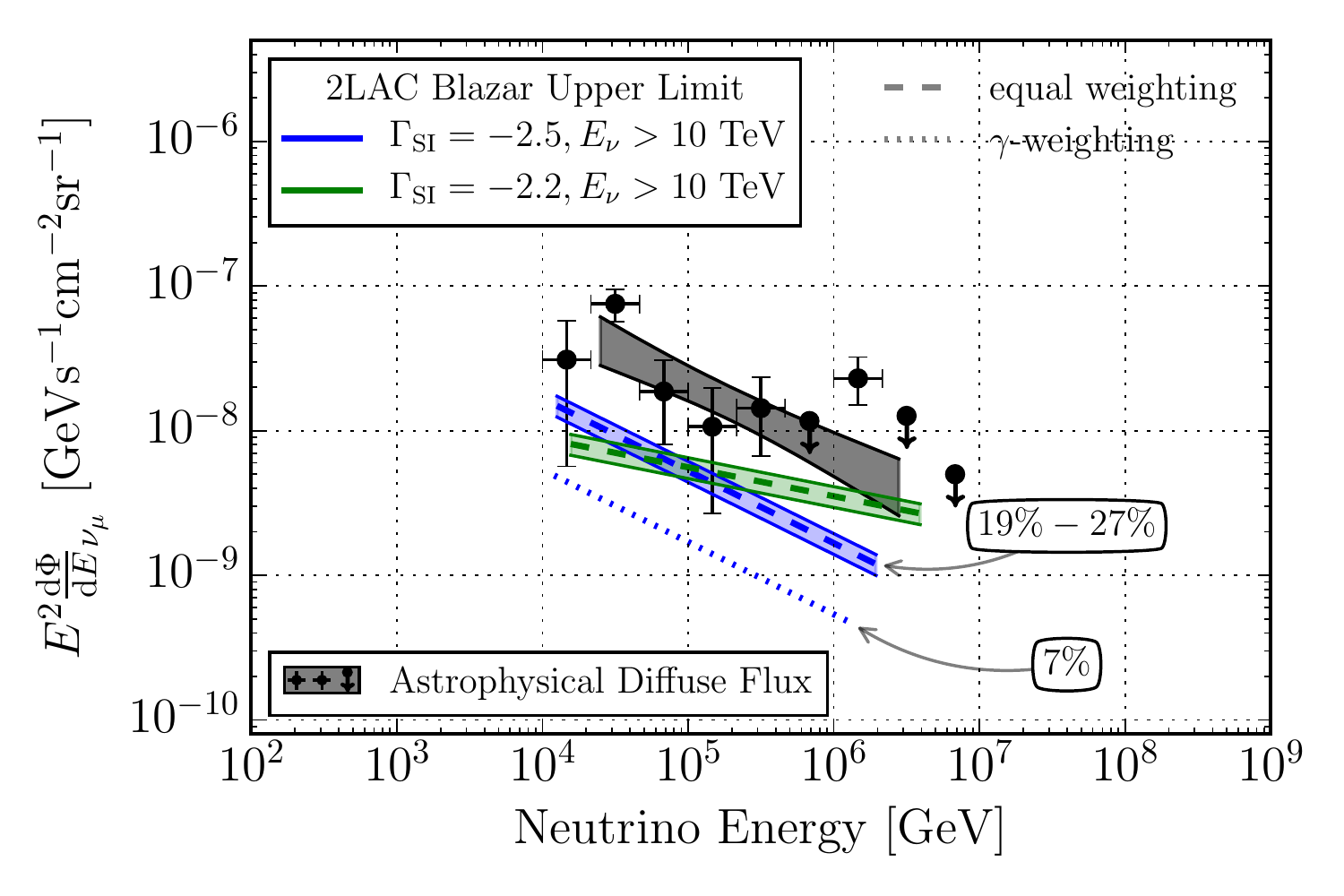}
\caption{$90 \% \ \mathrm{C.L.}$ flux upper limits for all 2LAC blazars in comparison to the observed astrophysical diffuse neutrino flux. The latest combined diffuse neutrino flux results from \citet{Aartsen2015a} are plotted as the best-fit power-law with spectral index $-2.5$ , and as a differential flux unfolding using $68 \%$ central and $90 \%$ U.L. confidence intervals.
The flux upper limit is shown using both weighting schemes for a power-law with spectral index $-2.5$ (blue). Percentages denote the fraction of the upper limit compared to the astrophysical best fit value. The equal-weighting upper limit for a flux with a harder spectral index of $-2.2$ is shown in green.}
\label{fig:blazar_contribution2}
\end{figure}

The astrophysical neutrino flux is observed
between 10~TeV and 2~PeV \citep{Aartsen2015a}. Its spectrum has been found to be compatible with a single power-law and a spectral index of $-2.5$ over most of this energy range. Accordingly, we use a power-law with the same spectral index and a minimum neutrino energy of 10~TeV for the signal injected into the simulated skymaps when calculating the upper limit for a direct comparison. Figure \ref{fig:blazar_contribution2} shows the flux upper limit for an $E^{-2.5}$ power-law spectrum starting at 10~TeV for both weighting schemes in comparison to the most recent global fit of the astrophysical diffuse neutrino flux, assuming an equal composition of flavors arriving at Earth.

The equal-weighting upper limit results in a maximally $19 \%$-$27 \%$ contribution of the total 2LAC blazar sample to the observed best fit value of the astrophysical neutrino
flux, including systematic uncertainties. This limit is independent of the detailed correlation between the $\gamma$-ray and neutrino flux from these sources. The only assumption is that the respective neutrino and $\gamma$-ray SCDs have similar shapes (see section \ref{section:upper_limits} for details on signal injection). We use the Fermi-LAT blazar SCD as published in \citet{Abdo2010a} as a template for sampling.
However, we find that even if the shape of the SCD differs from this template, the upper limit still holds and is robust. In appendix \ref{appendix:scd_dependence} we discuss the effect of different SCD shapes and discuss how the combination with existing point source constraints \citep{Aartsen2015b} leads to a nearly SCD-independent result, since a point source analysis and a stacking search with equal weights effectively trace opposite parts of the available parameter space for the $dN/dS$ distribution.

In case we assume a proportionality between the $\gamma$-ray and neutrino luminosities of the sources, the $\gamma$-weighting limit constrains the maximal flux contribution of all 2LAC blazars to $7 \%$ of the observed neutrino flux in the full 10 TeV to 2 PeV range. Since the blazars resolved in the 2LAC account for $70 \ \%$ of the total $\gamma$-ray emission from all GeV blazars \citep{Ajello2015}
this further implies that at most $10 \% $ of the astrophysical neutrino flux stems from all GeV blazars extrapolated to the whole universe, again in the full 10 TeV to 2 PeV range and assuming the $\gamma$-weighting is an appropriate weighting assumption.
Table \ref{table:relative_contribution} summarizes the maximal contributions for all populations, including the $\gamma$-weighting result scaled to the total respective total population of sources in the observable universe.

It is interesting to compare these numbers directly to the $\gamma$-ray sector. \citet{Ajello2015} show that GeV blazars ($100 \ \mathrm{MeV}-100 \ \mathrm{GeV}$) contribute approximately $50 \%$ to the extragalactic gamma-ray background (\textbf{EGB}). The resolved 1FGL \citep{Abdo2010b} blazar component in particular contributes around $35 \%$. This estimate should be rather similar for the 2LAC blazars studied here, which are defined based on the more recent 2FGL catalogue \citep{Nolan2012} (see appendix \ref{appendix:correction_factor} for a discussion). The 2LAC blazar contribution to the astrophysical neutrino flux is therefore by at least a factor $0.75$ smaller than the corresponding extragalactic contribution in the $\gamma$-regime. The difference of this contribution between the two sectors becomes substantial ($7 \%$ maximally allowed contribution for neutrinos versus $35 \%$ for $\gamma$-rays) if one assumes a $\gamma$/$\nu$-correlation.

\begin{table}
	\centering
	\begin{tabular}{ c||c|c|c }
		\hline
		\hline
		\multirow{2}[2]{*}{Population} & \multicolumn{3}{c}{ \rule{0pt}{2.5ex} weighting scheme  } \\ & equal & $\gamma$ & $\gamma$ (extrapol.)  \\
		\hline
		\hline
		all 2LAC blazars &  $19 \%-27 \%$ & $7 \%$  & $10 \%$  \\
		\hline
		FSRQs & $5 \%-17 \%$  &  $5 \%$ &  $7 \%$  \\
		\hline
		LSPs  & $6 \%-15 \%$   &  $5 \%$   & $7 \%$   \\
		\hline
		ISP/HSPs  & $9 \%-15 \%$   &  $5 \%$   & $7 \%$   \\
		\hline
		LSP-BL Lacs  & $3 \%-13 \%$   &   $6 \%$  & $9 \%$   \\
		\hline
	\end{tabular}
	\caption{Maximal contributions to the best-fit diffuse flux from \citet{Aartsen2015a} assuming equipartition of neutrino flavours.
		The equal-weighting case shows this maximal contribution for the $90 \%$ central outcomes of potential dN/dS realizations.  The last column shows the maximal contribution of the integrated emission from the total parent population in the observable universe exploiting the $\gamma$-ray completeness of the 2LAC blazars (see appendix \ref{appendix:correction_factor}).   
	}
	\label{table:relative_contribution}
\end{table}

Figure \ref{fig:blazar_contribution2} also shows the equal-weighting constraint for a harder neutrino spectrum with a spectral index of $-2.2$. This harder spectral index is about 3 standard deviations away from the best fit value derived in \citet{Aartsen2015a}, and can be used as an extremal case given the current observations. The comparison of this upper limit with the hard end of the "butterfly" shows that even in this case less than half of the bulk emission can originate in the 2LAC blazars with minimal assumptions about the relative neutrino emission strengths. Due to the low-count status of the data, we omit multi power-law spectra tests at this point. However, one can estimate the constraints for more complicated models using figure \ref{fig:energy_interval} in appendix \ref{appendix:energy_range}, which shows the energy range for a given spectrum that contributes the dominant fraction to the sensitivity. The sensitivity for a possible two-component model, for example, having a soft component at TeV energies and a hard component in the PeV range, would be dominated by the soft regime, as the "ratio function" (see appendix \ref{appendix:energy_range}, figure \ref{fig:energy_interval}) by the hard component above a PeV is negligible. In such a scenario we expect the constraint to be rather similar to our result from the simple power-law test with spectral index $-2.5$.

\subsection{Upper limits on models for diffuse neutrino emission}
\label{section:upper_limits_models}

For the experimental constraints on existing theoretical calculations, we only considered models for the diffuse emission from blazar populations, not predictions for specific objects. These include the calculations by 
\citet{Mannheim1995}, \citet{Halzen1997} and \citet{Protheroe1997} for generic blazars, the calculations by \citet{Becker2005} and \citet{Murase2014} for FSRQs and calculations by \citet{Muecke2003},\citet{Tavecchio2014},\citet{Tavecchio2015} and \citet{Padovani2015} for BL\,Lacs. 

The upper limits in this section are calculated using the $\gamma$-weighting scheme and therefore assume a correlation between the neutrino flux and the measured $\gamma$-ray energy flux. This allows us to account for the fraction of the neutrino emission that arises from the blazars not detected in $\gamma$-rays.
The fraction of $\gamma$-ray emission from resolved 2LAC blazars in general (including BL\,Lacs), and of FSRQs in particular, 
is about $70 \%$ \citep{Ajello2015,Ajello2012}. Therefore, the flux upper limits for the entire population are a factor $1/0.7 \approx 1.43$ weaker than those derived for the quasi-diffuse flux of the 2LAC blazars. See appendix \ref{appendix:correction_factor} for more details on this factor.

\begin{table*}
\centering
\begin{threeparttable}
\begin{tabular}{c|cc|c}
\hline
Type & \multicolumn{2}{c|}{Model} & MRF
\\
\hline
\hline
\multirow{4}[4]{*}{Generic blazars} & \multirow{2}[0]{*}{\citep{Mannheim1995}} & (A) & $1.30$ \\ [0.1cm]
 &  & (B) & $<0.1$ \\ 
 \cline{2-4}
&  \citep{Halzen1997} & & $<0.1$ \\ 
\cline{2-4}
&  \citep{Protheroe1997} & & $<0.1$  \\ 
\hline
\multirow{5}[5]{*}{FSRQs} & \citep{Becker2005} &  & $2.28$ \\
\cline{2-4}
 & \multirow{4}[5]{*}{\citep{Murase2014}} & $\Gamma_{\mathrm{SI}}=-2.0$ (BLR)& $ \xi_{\mathrm{CR}}<12$ \\
 &  & $\Gamma_{\mathrm{SI}}=-2.0$ (blazar) & $ \xi_{\mathrm{CR}}<21$ \\
 &  & $\Gamma_{\mathrm{SI}}=-2.3$ (BLR) & $\xi_{\mathrm{CR}}<153 $ \\
  &  & $\Gamma_{\mathrm{SI}}=-2.3$ (blazar) & $ \xi_{\mathrm{CR}}<241$ \\
\hline
\multirow{3}[8]{*}{BL\,Lacs} &  \multirow{2}[0]{*}{\citep{Muecke2003}} & HSP (optimistic) & $76.29$ \\
 &   & LSP (optimistic) & $5.78$ \\
 \cline{2-4}
&  \multirow{2}[0]{*}{ \begin{tabular}{c} \citep{Tavecchio2014} \\  \end{tabular}  }  & HSP-dominated (1) & $1.06$ \\
&  \tnote{a} & HSP-dominated (2) & $0.35$ \\
& \citep{Tavecchio2015}  & LSP-dominated & $0.21$ \\ 
\cline{2-4}
&  \citep{Padovani2015} & HSP (baseline) & $0.75$  \\ 
\hline
 \end{tabular}
 \begin{tablenotes}
   \item[a] Predictions from \citet{Tavecchio2014,Tavecchio2015} enhanced by a factor 3 in correspondence with the authors.
 \end{tablenotes}
 \end{threeparttable}
\caption{Summary of constraints and model rejection factors for the diffuse neutrino flux predictions from blazar populations. The values include a correction factor for unresolved sources (see appendix \ref{appendix:correction_factor}) and systematic uncertainties. For models involving a range of flux predictions we calculate the MRF with respect to the lower flux of the optimistic templates \citep{Muecke2003} or constraints on baryon to photon luminosity ratios $\xi_{\mathrm{CR}}$ \citep{Murase2014}. }
\label{table:model_rejection_factors}
\end{table*}

\begin{figure*}
	\centering
	\begin{tabular}[b]{@{}p{0.45\textwidth}@{}}
    \centering\includegraphics[width=.45\textwidth]{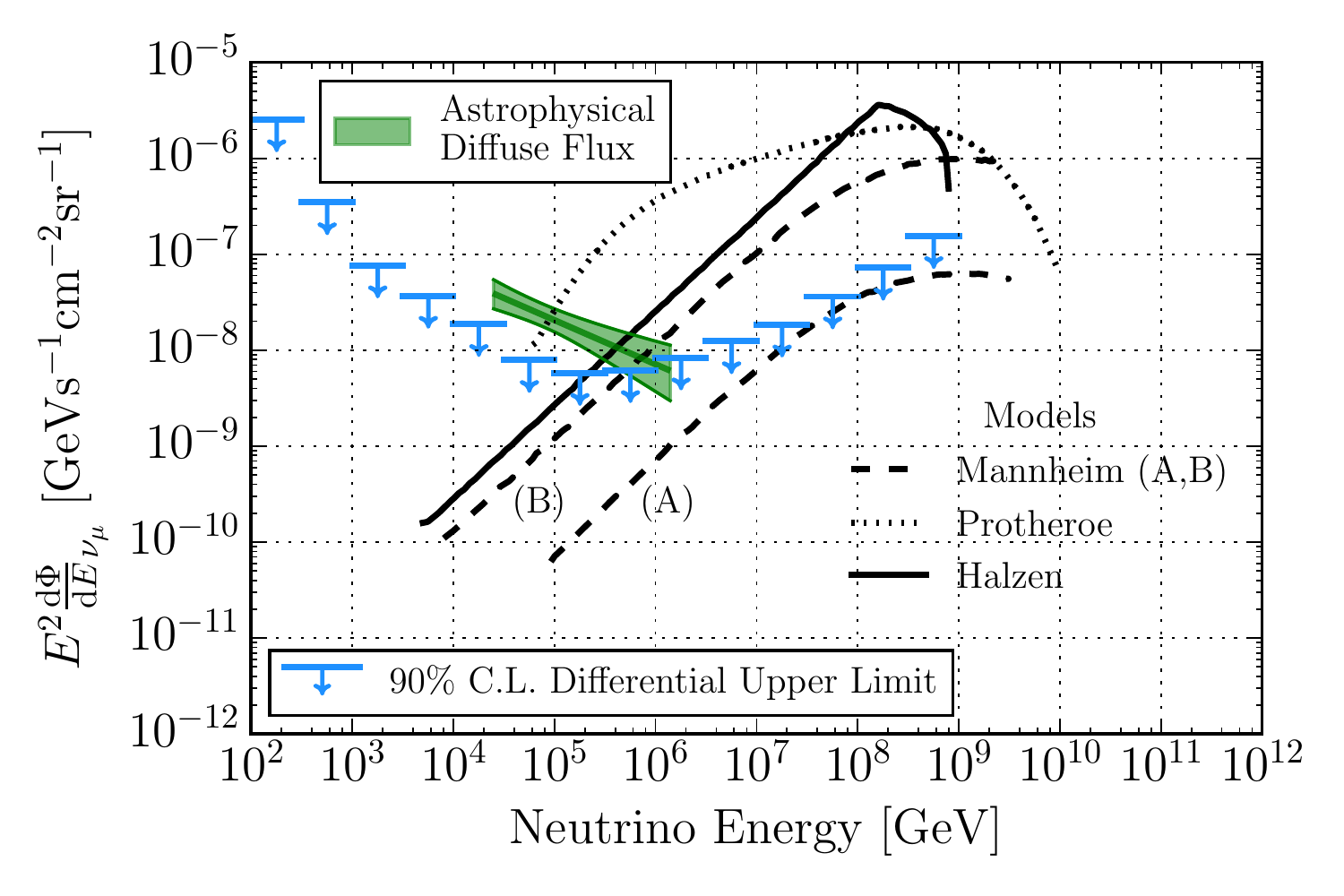} \\
    \centering\small (a) generic blazars
  	\end{tabular}%
  	\quad
  	\begin{tabular}[b]{@{}p{0.45\textwidth}@{}}
    \centering\includegraphics[width=.45\textwidth]{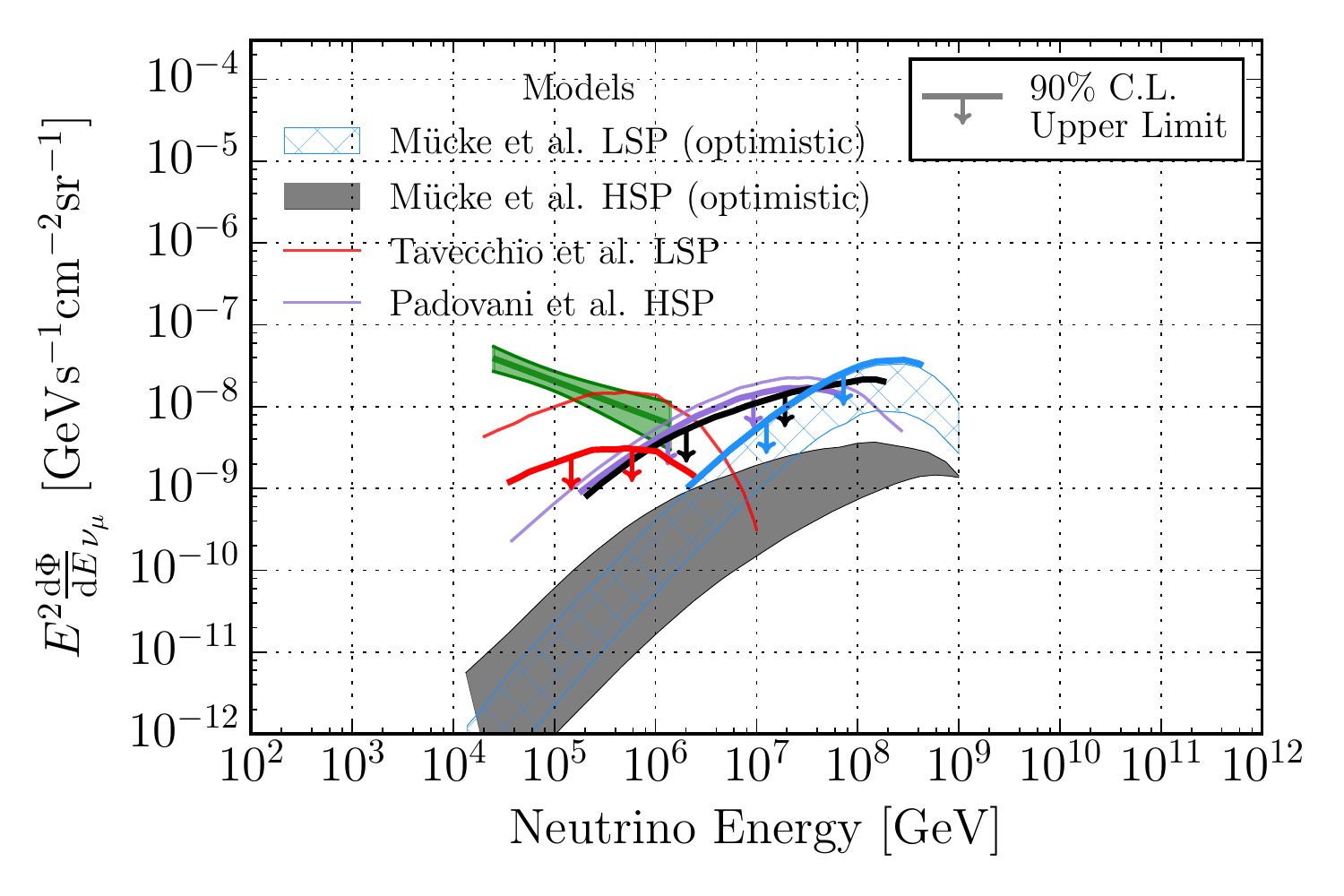} \\
    \centering\small (b) BL\,Lacs
  	\end{tabular} \\
  	\begin{tabular}[b]{@{}p{0.45\textwidth}@{}}
    \centering\includegraphics[width=.45\textwidth]{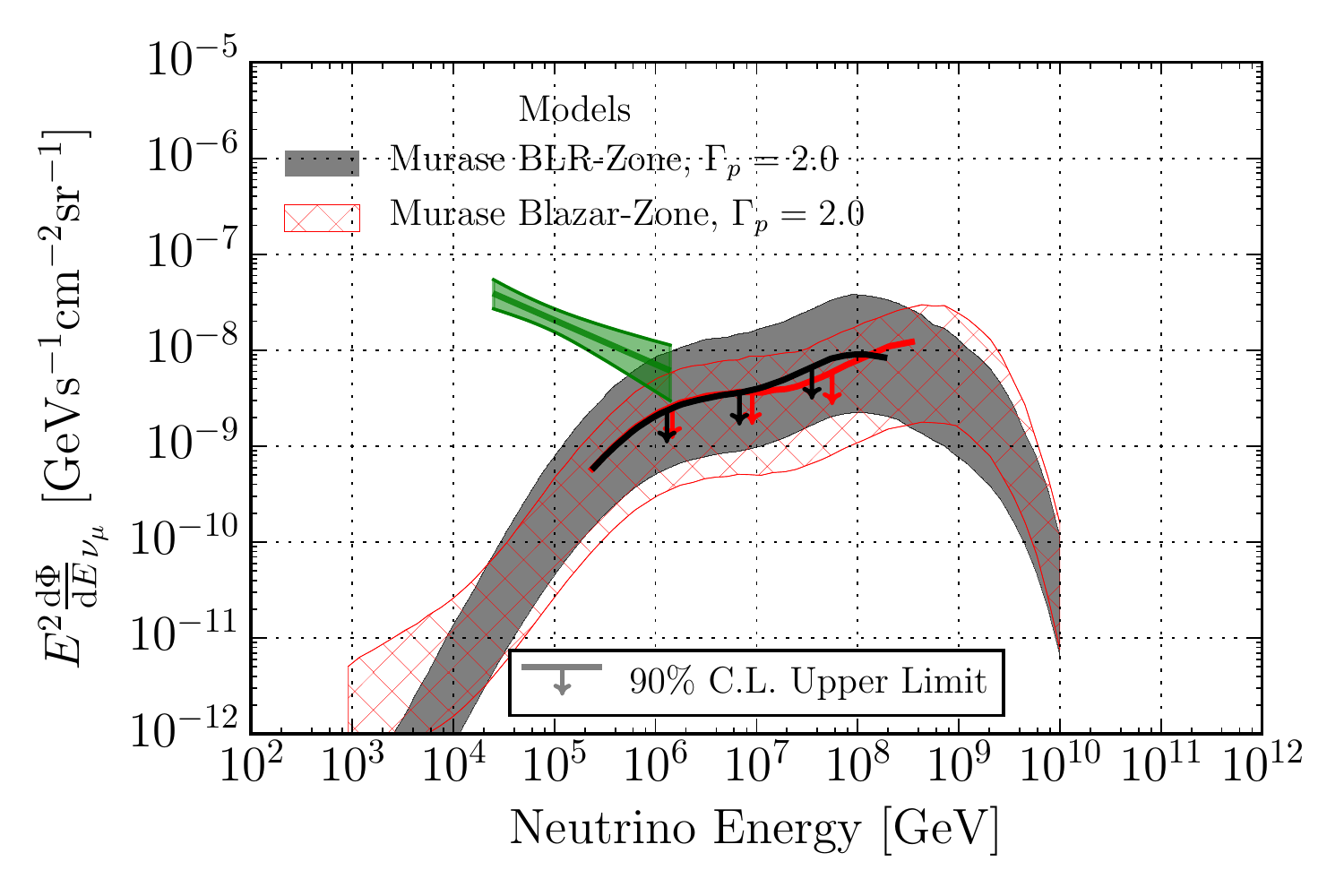} \\
    \centering\small (c) FSRQs - 1
  	\end{tabular}%
  	\quad
  	\begin{tabular}[b]{@{}p{0.45\textwidth}@{}}
    \centering\includegraphics[width=.45\textwidth]{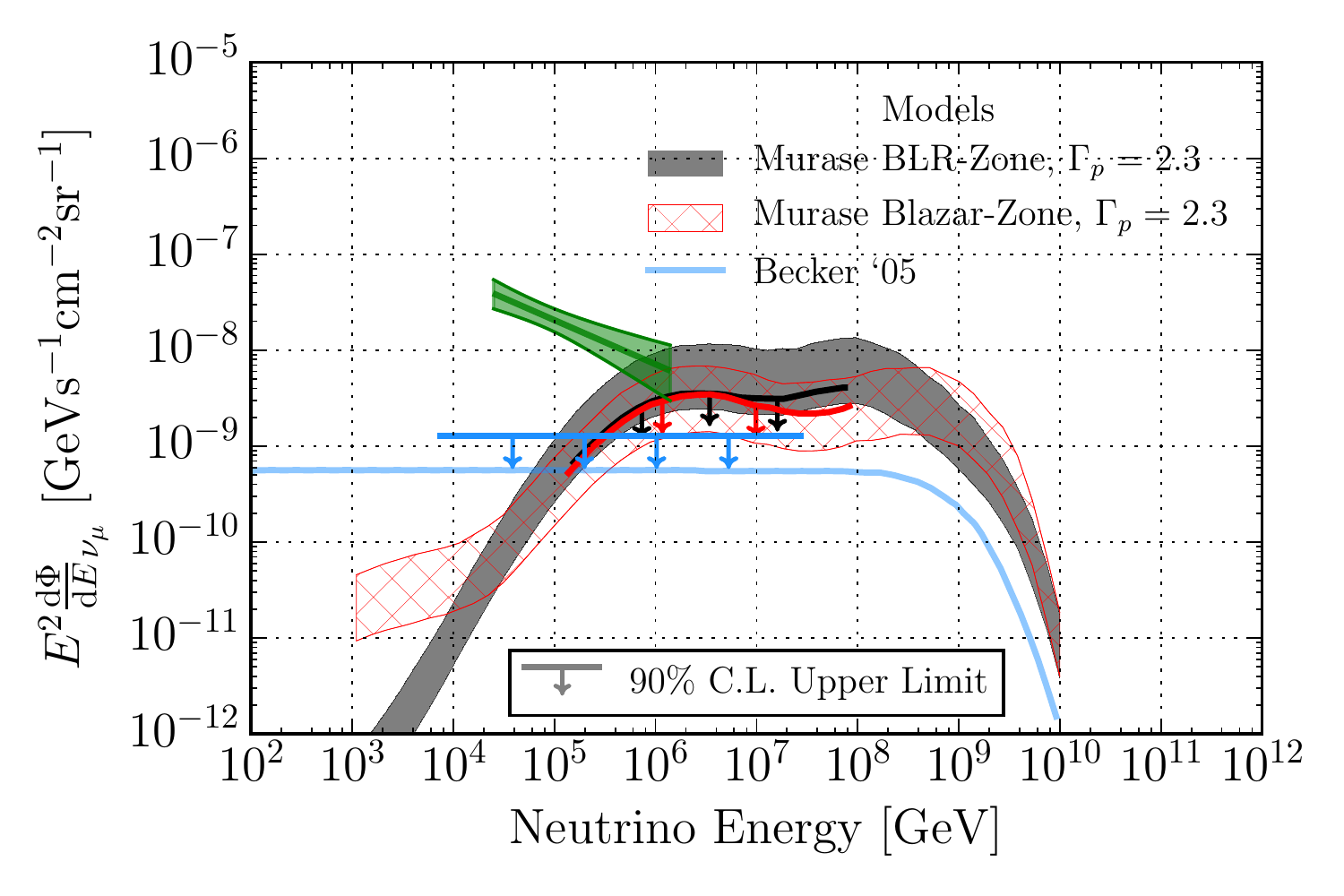} \\
    \centering\small (d) FSRQs - 2
  	\end{tabular}
  	\caption{$90 \% \ \mathrm{C.L.}$ upper limits on the ($\nu_\mu + \overline{\nu}_\mu$)-flux for models of the neutrino emission from (a) generic blazars \citep{Mannheim1995,Halzen1997,Protheroe1997}, (b) BL\,Lacs \citep{Muecke2003,Tavecchio2015,Padovani2015} and (c)+(d) FSRQs \citep{Becker2005,Murase2014}. The upper limits include a correction factor that takes into account the flux from unresolved sources (see appendix \ref{appendix:correction_factor}) and systematic uncertainties. The astrophysical diffuse neutrino flux measurement \citep{Aartsen2015a} is shown in green for comparison.}
  	\label{fig:neutrino_model_uls}
\end{figure*}

Table \ref{table:model_rejection_factors} summarizes model rejection 
factors \citep{Hill2003}\footnote{The flux upper limit divided by the 
flux predicted in the model.} for all considered models.
Many of these models can be constrained by this analysis. Figure \ref{fig:neutrino_model_uls} a)-d) visualizes the flux upper limits in comparison to the neutrino flux predictions. 

In the early models (before the year 2000) the neutrino flux per source is calculated as 
being directly proportional to the $\gamma$-ray flux in the energy range $E_{\gamma}>100 \ \mathrm{MeV}$\citep{Mannheim1995}
(A),  $E_{\gamma}>1 \ \mathrm{MeV}$ \citep{Mannheim1995} (B), 
$20 \ \mathrm{MeV} < E_{\gamma} < 30 \ \mathrm{GeV}$ \citep{Halzen1997} 
and $E_{\gamma}>100 \ \mathrm{MeV}$ \citep{Protheroe1997}. The $\gamma$-weighting 
scheme is therefore almost implicit in all these calculations, although the energy ranges vary slightly from the $100 \ \mathrm{MeV} - 100 \ \mathrm{GeV}$ energy range used for the $\gamma$-weighting. 

From the newer models, only \citet{Padovani2015} uses a direct proportionality between neutrino and $\gamma$-ray flux (for $E_{\gamma}>10 \ \mathrm{GeV}$), where the proportionality factor encodes a possible leptonic contribution. In all other publications a direct correlation to $\gamma$-rays is not used for the neutrino flux calculation. 
Since all these models assume that p/$\gamma$-interactions dominate the neutrino production, the resulting neutrino fluxes are calculated via the luminosity in the target photon fields. In \citet{Becker2005} 
the neutrino flux is proportional to the target radio flux which in turn is connected to the disk luminosity 
via the model from \citet{Falcke1995}. In \citet{Muecke2003} it is directly proportional to the radiation of the synchrotron peak. In \citet{Murase2014} the neutrino flux is 
connected to the x-ray luminosity, which in turn is proportional to the luminosity in various target photon fields. In \citet{Tavecchio2014} the neutrino luminosity is calculated using target photon fields from the inner jet ``spine-layer''. 
However, a correlation to the $\gamma$-ray flux in these latter models may still exist, even in the case that leptonic $\gamma$-ray contributions dominate. This is mentioned in \citet{Murase2014}, which explicitly predicts the strongest $\gamma$-ray emitters to be also the strongest neutrino emitters, even though the model contains leptonically produced $\gamma$-ray emission. It should be noted that an independent IceCube analysis studying the all-flavor diffuse neutrino flux at PeV energies and beyond \citep{Aartsen2016b} recently also put strong constraints on some of the flux predictions discussed in this section.

\section{Summary and outlook}
\label{section:discussion}

In this paper, we have analyzed all 862 Fermi-LAT 2LAC blazars and 4 spectrally selected sub populations 
via an unbinned likelihood stacking approach for a cumulative neutrino excess from the given blazar directions. The study uses 3 years of IceCube data (2009-2012) amounting to a total of around $340000$ muon-track events. 

Each of the 5 populations were analyzed with two weighting schemes which encode the assumptions 
about the relative neutrino flux from each source in a given population.
The first weighting scheme uses the energy flux observed in  $\gamma$-rays as weights, 
the second scheme gives each source the same weight. This resulted in a total of 10 statistical tests which were in turn analyzed in two different ways. 
The first is an ``integral'' test, in which a power-law flux with a variable spectral 
index is fitted over the full energy range that IceCube is sensitive to. 
The second is a differential analysis, in which 14 energy segments between $10^2 \ \mathrm{GeV}$ and $10^9 \ \mathrm{GeV}$, each spanning half a decade
in energy, are fit independently  with a constant spectral index of $-2$.

Nine from ten integral tests show over-fluctuations, but none of them are significant. The largest overfluctuation, a $6\%$ p-value, is observed for all 862 2LAC blazars combined using the model-independent equal-weighting scheme. The differential test for all 2LAC blazars using equal source weighting reveals that the excess appears in the 5-10 TeV region with a local p-value of $2.6 \sigma$. No correction for testing multiple hypotheses is applied, since even without a trial correction this excess cannot be considered significant.

Given the null results we then calculated flux upper limits.
The two most important results of this paper are:

\begin{enumerate}

\item{We calculated a flux upper limit for a power-law spectrum starting at $10 \ \mathrm{TeV}$ with a spectral index of $-2.5$ for all 2LAC blazars. We compared this upper limit to the diffuse 
astrophysical neutrino flux observed by IceCube \citep{Aartsen2015a}. We found that the maximal contribution from all 
2LAC blazars in the energy range between 10 TeV and 2 PeV is at most $27 \%$, including systematic effects and with minimal assumptions about the neutrino/$\gamma$-ray correlation in each source. Changing the spectral index of the tested flux to $-2.2$, a value allowed at about 3 standard deviations given the current global fit result \citep{Aartsen2015a}, weakens this constraint by about a factor of two.
If we assume for each source a similar proportionality between the $\gamma$-ray luminosity in the 2LAC energy range and the neutrino luminosity, we can extend the constraint to the parent population of all GeV blazars in the observable universe. The corresponding maximal contribution from all GeV blazars is then around $10 \%$, or $5-10 \%$ from the other blazar sub-populations. In each case we use the same power-law assumption as before in order to compare to the observed flux. For FSRQs our analysis allows for a $7 \%$ contribution to the diffuse flux, which is in general agreement with a result found by \citet{Wang2015} who independently estimated that FSRQs do not contribute more than $10 \%$ to the diffuse flux using our earlier small-sample stacking result for 33 FSRQs \citep{Aartsen2014}.
}

\item{
We calculated upper limits using the $\gamma$-weighting scheme for 15 models of the diffuse neutrino emission from blazar populations found in the literature. For most of these models, the upper limit constrains the model prediction, for some of them by more than an order of magnitude. The implicit assumption in all these upper limits is a proportionality between the source-by-source $\gamma$-ray luminosity in the 2LAC energy range and its corresponding neutrino luminosity. All models published before the year 2000, and the model by \citet{Padovani2015} implicitly contain this assumption, although some of their energy ranges differ from the exact energy range in the 2LAC catalogue. Even for the other models the proportionality assumption may still hold, as indicated by \citet{Murase2014}. 
}

\end{enumerate}

\citet{Kadler2016} recently claimed a $5 \%$ chance probability for a PeV IceCube event to have originated close to blazar \texttt{PKS B1424-418} during a high-fluence state. While $5 \%$ is not yet statistical evidence, our results do not contradict such single PeV-event associations, especially since a dominant fraction of the sensitivity of our analysis comes from the sub-PeV energy range. The same authors also show that the measured all-sky PeV neutrino flux can not be compatible with an origin in a pure FSRQ population that has a peaked spectrum around PeV energies, as it would over-predict the number of observed events. Instead, one has to invoke additional assumptions, for example a certain contribution from BL Lacs, leptonic contributions to the SED, or a spectral broadening of the arriving neutrino flux down to TeV energies due to Doppler shifts from the jets and the intrinsic redshift distribution of the blazars. Our results suggest that the last assumption, a spectrum broadening down to TeV energies, only works if the resulting power-law spectral index is harder than around $-2.2$, as the flux is otherwise in tension with our $\gamma$-weighting upper limit. A hard PeV spectrum is interestingly also seen by a recent IceCube analysis \citep{Aartsen2016} that probes the PeV range with muon neutrinos. Regardless of these speculations, the existing sub-PeV data requires an explanation beyond the 2LAC sample from a yet unidentified galactic or extra-galactic source class.

Our results do not provide a solution to explain the bulk emission of the astrophysical diffuse 
neutrinos, but they provide robust constraints that might help to construct the global picture. 
Recently, \citet{Murase2015} argued that current observations favour sources that are opaque to $\gamma$-rays. This would for example be expected in the cores of AGN. Our findings on the 2LAC blazars mostly probe the emission from relativistically beamed AGN jets and are in line with these expectations. We also do not constrain neutrinos from blazar classes that are not part of the 2LAC catalogue, for example extreme HSP objects. These sources might emit up to $30 \%$ of the diffuse flux \citep{Padovani2016}, and studies in this direction with other catalogues are in progress.  

While the slight excess in the 5-10 TeV region is not yet significant, further observations by IceCube may clarify if we see an emerging soft signal or just a statistical fluctuation.

\acknowledgments

We acknowledge the support from the following agencies: U.S. National Science Foundation-Office of Polar Programs, U.S. National Science Foundation-Physics Division, University of Wisconsin Alumni Research Foundation, the Grid Laboratory Of Wisconsin (GLOW) grid infrastructure at the University of Wisconsin - Madison, the Open Science Grid (OSG) grid infrastructure; U.S. Department of Energy, and National Energy Research Scientific Computing Center, the Louisiana Optical Network Initiative (LONI) grid computing resources; Natural Sciences and Engineering Research Council of Canada, WestGrid and Compute/Calcul Canada; Swedish Research Council, Swedish Polar Research Secretariat, Swedish National Infrastructure for Computing (SNIC), and Knut and Alice Wallenberg Foundation, Sweden; German Ministry for Education and Research (BMBF), Deutsche Forschungsgemeinschaft (DFG), Helmholtz Alliance for Astroparticle Physics (HAP), Research Department of Plasmas with Complex Interactions (Bochum), Germany; Fund for Scientific Research (FNRS-FWO), FWO Odysseus programme, Flanders Institute to encourage scientific and technological research in industry (IWT), Belgian Federal Science Policy Office (Belspo); University of Oxford, United Kingdom; Marsden Fund, New Zealand; Australian Research Council; Japan Society for Promotion of Science (JSPS); the Swiss National Science Foundation (SNSF), Switzerland; National Research Foundation of Korea (NRF); Villum Fonden, Danish National Research Foundation (DNRF), Denmark

\appendix

\section{Dependence of flux upper limits on the source count distribution sampling}
\label{appendix:scd_dependence}

\begin{figure*}
\epsscale{1.0}
\plotone{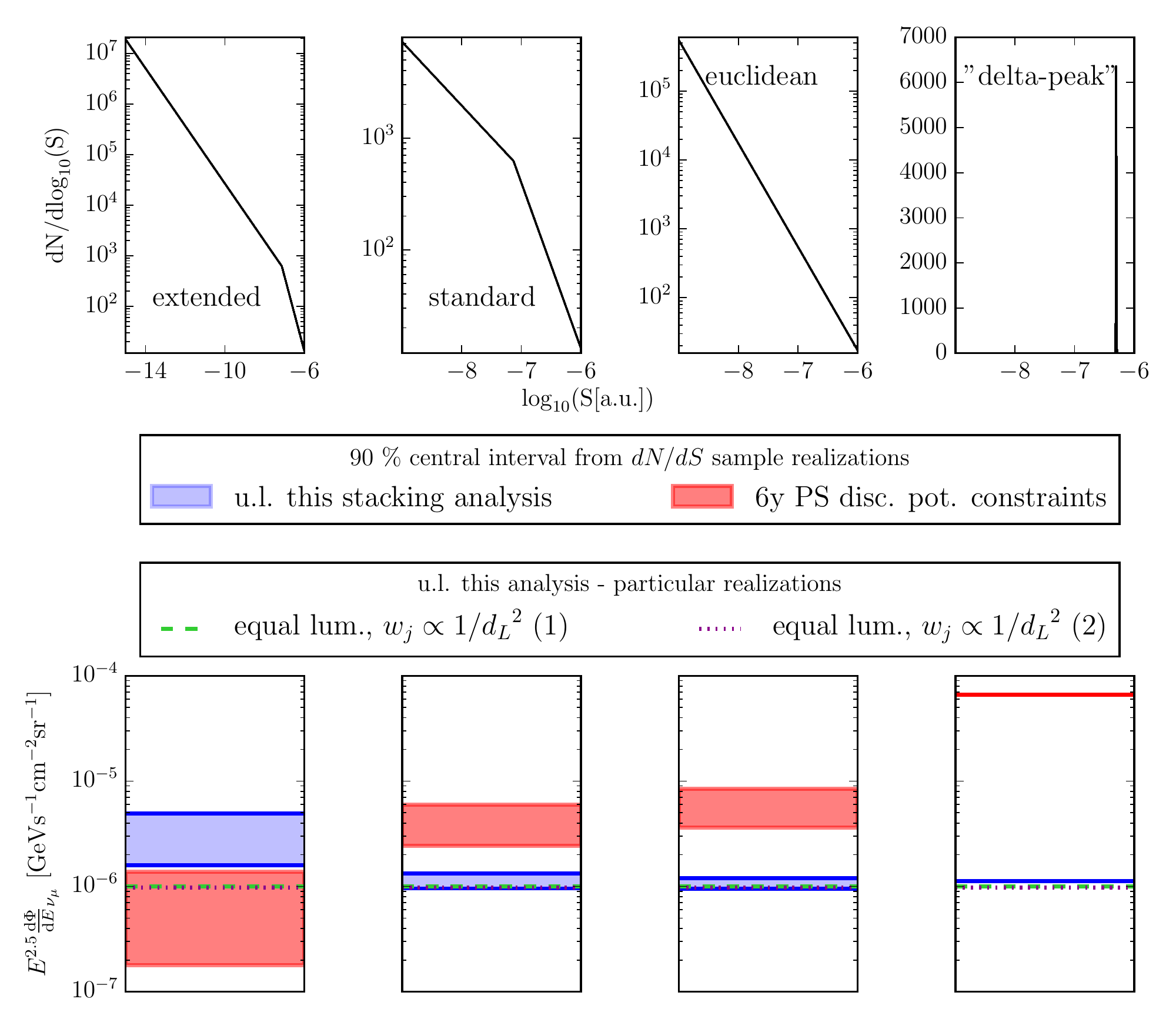}
\caption{Comparison of equal-weighting upper limits for different SCDs which are used to sample relative source injection weights, shown for the population of all 2LAC blazars. The upper row shows the SCDs and the lower row the respective constraints for an $E^{-2.5}$ flux starting at $10$ TeV. The source flux $S$ on the x-axis is shown in arbitrary units, since only the relative neutrino flux counts, but orients itself by the integrated $\gamma$-ray flux from \citet{Abdo2010a}. The light blue band marks the $90 \%$ central interval of upper limit outcomes for random samplings of the given SCD, the light red band marks the constraints from the 6-year PS search for similar random samplings. Two specific realizations modeling equal intrinsic luminosity, taking into account the luminosity distance and randomly drawn redshifts for missing-redshift BL\,Lacs, are shown in green and magenta.  
}
\label{fig:equal_weighting_construction}
\end{figure*}

The equal-weighting limits use source count distributions to model the neutrino injection. The SCD serves as a PDF template from which relative neutrino injection weights are drawn. 
Depending on the shape of the SCD, and the range of flux values in which the SCD is being sampled, the resulting central neutrino upper limit value shifts and the range of calculated flux upper limits broadens. This is illustrated in figure \ref{fig:equal_weighting_construction}, which shows the upper limits derived from ensemble simulations drawn from four different SCDs. It also contains standard 6-year point source constraints \citep{Aartsen2015b} for similar flux realizations.

All examples are shown for a population of 862 objects, the size of the ``All 2LAC blazars'' sample.
In the left panel (``extended''), the SCD template is the measured blazar $\gamma$-ray SCD from \citet{Abdo2010a} and is extrapolated five orders of magnitude to lower flux values. The minimum flux value is arbitrarily chosen, but it is small enough such that the distribution is a scale-free power-law
and an extension towards even smaller flux values does not make a difference in terms of average sample outcomes. 
In the second panel (``standard''),the SCD is exactly the Fermi-LAT blazar SCD from \citet{Abdo2010a} and spans three orders of magnitude in flux. In the third panel the SCD is of Euclidean form and extended over a flux range such that the cumulative SCD equals to the number count of the ``standard'' distribution in the second panel.
In the fourth panel the SCD is a delta distribution, which gives an equal weight to each source, i.e. the assumption that is used in the weighting of the PDF for the statistical test.
The lower row displays the respective $90 \%$ central interval for upper limit outcomes from this analysis and constraints from 6-year single point source search discovery potentials. 

As we sample the relative source contributions from a growing flux range, corresponding to the column order $4 \rightarrow 3 \rightarrow 2 \rightarrow 1$, flux upper limit variations increase and the stacking analysis constraint weakens. At the same time, the constraints from the single point source search become stronger with a single source more and more dominating the total population.

The first and fourth columns correspond to limiting cases, neither of which are appropriate to use for this analysis, but are just shown to illustrate the general behavior of the procedure. The delta-peak SCD (4th column) is unphysical, since it corresponds to an equal flux per source. The extrapolated SCD (1st column) yields an extreme spread of signal contributions, which roughly corresponds to a random draw from the entire population in the universe, assuming that the faint end corresponds to the weakest blazars that can in principle be detected. Since the random draw mostly exists of sources from the faint-flux end, it corresponds to a situation where the neutrino flux is anti-proportional to the $\gamma$-ray flux, which is unphysical. All results in this paper make use of the 2nd column SCD. The cross-over of point source constraints and constraints from this stacking analysis is reached for realizations drawn from a $dN/dS$ distribution that lies in between the SCD from the 1st and 2nd column. For illustrative purposes we also include a particular flux scenario with equal intrinsic luminosity, where the neutrino flux per source is proportional to $1/{d_L}^2$ \footnote{Using standard $\Lambda$CDM cosmological parameters as determined by the Planck mission \citep{Planck2015}.}. Missing redshifts for BL\,Lacs are drawn randomly from the BL\,Lac distribution where the redshifts are known, taking into account the synchroton peak information. The results for two such realizations of missing redshift sources lie in the range of SCD draws from the 2nd column. Since the two realizations do not differ significantly, we conclude that the resulting estimate is robust, even though some of the BL\,Lac redshifts are unknown.  
 
\section{Determination of energy range for upper limits}
\label{appendix:energy_range} 
 
\begin{figure*}
\epsscale{0.7}
\plotone{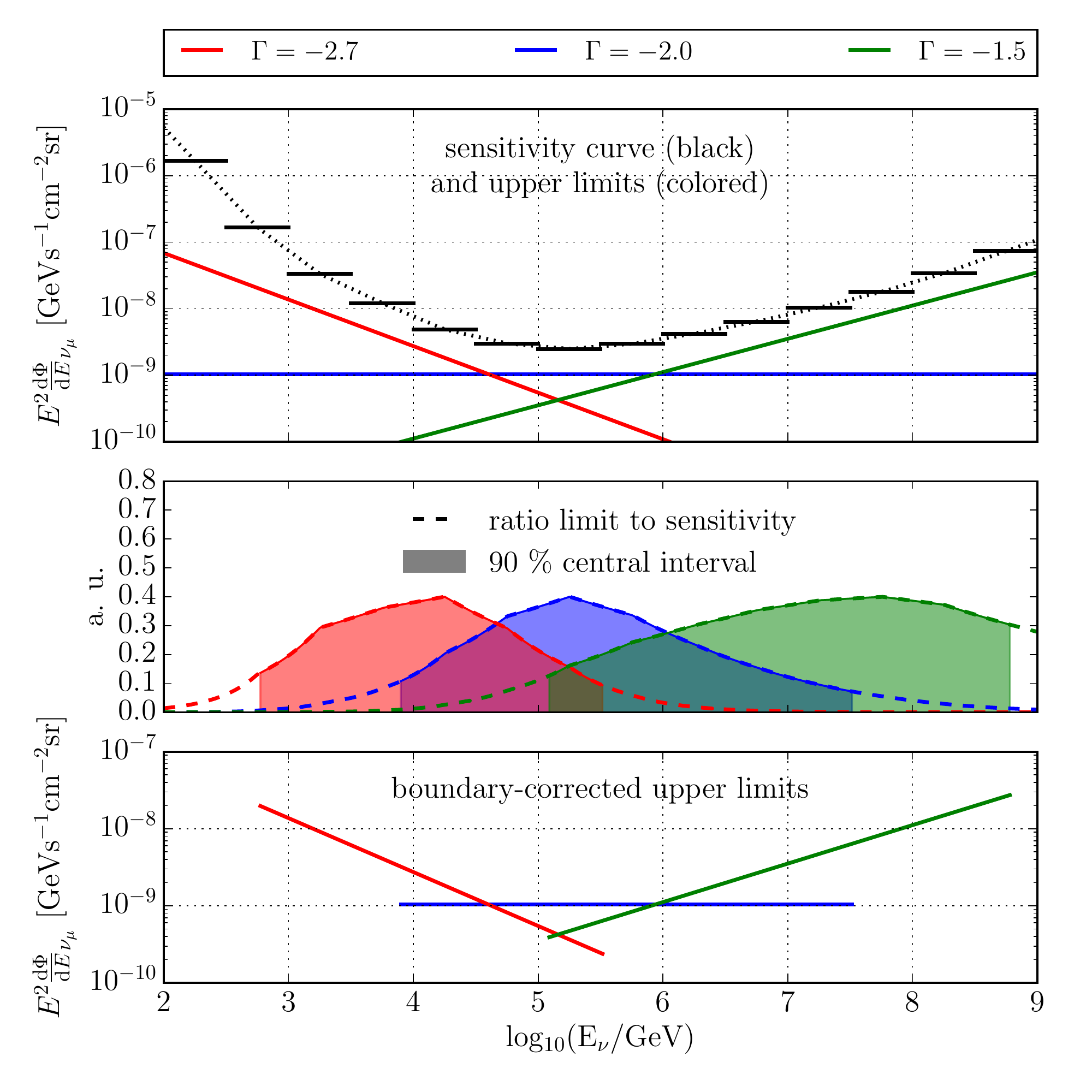}
\caption{Determination of the energy range that contributes $90\%$ to the total sensitivity of 
IceCube for a neutrino flux of a given spectrum. The construction is shown for the total
2LAC-blazar population using the $\gamma$-energy flux weighting scheme for three 
power-law spectra with spectral indices $-1.5$, $-2.0$ and $-2.7$.}
\label{fig:energy_interval}
\end{figure*}

We determine the energy range for which the upper limit is supported by IceCube data based on 
the differential sensitivity. The procedure is illustrated in figure \ref{fig:energy_interval} for three different generic power-law spectra using the $\gamma$-weighting scheme. The ratio of the differential sensitivity curve with a convex energy spectrum generally forms a function with a single maximum that falls off towards the sides.
The central interval enclosing $90 \%$ of the area under the ratio function is used to define the energy range that contributes $90 \%$ to the sensitivity for a given energy spectrum. This area-sensitivity relation has been checked empirically. The methodology is an extension to a previous method, e.g. used in \citet{Aartsen2014}, which only uses the $90 \ \%$ central interval of signal events and thereby neglects the background rate.  

%

\section{Upper limit correction factors for unresolved sources}
\label{appendix:correction_factor}

The $\gamma$-weighting scheme implicitly assumes a proportionality between neutrino and $\gamma$-ray luminosities. In this case, the fraction of the total neutrino flux of the population
that originates from the source resolved in $\gamma$-rays is equal to the fraction of the total $\gamma$-ray flux that originates from the resolved sources. It has been estimated that $70 \%$ of the total diffuse $\gamma$-ray flux from blazars between $100 \ \mathrm{MeV}$ and $100 \mathrm{GeV}$ has been resolved in 1FGL blazars at high galactic latitudes $|b| > 15^\circ$ \citep{Ajello2015}, and similarly for 1FGL FSRQs \citep{Ajello2012} in particular. The 2LAC catalogue used in this work contains blazars at galactic latitudes $|b| > 10^\circ$. At the extra galactic latitudes ($10^\circ < |b| < 15^\circ$) the detection efficiency might be worse. However, even if it unrealistically sharply drops to zero, one can estimate that the total resolved fraction does only shrink by an amount that is proportional to the ratio of the $10^\circ < |b| < 15^\circ$ sky fraction ($\approx 8 \%$ of the total sky) with respect to the rest ($\approx 74 \%$ of the total sky ), i.e. to $\frac{0.08 \cdot 0 + 0.74 \cdot 70 \%}{0.82} \approx 63 \%$. Since this estimate is conservative, is still within the error on the quoted $70 \%$ value, and since the 2LAC sample is based on the more sensitive 2FGL catalogue, we conclude that the $70 \%$ completeness is a reasonable estimate to choose for the 2LAC sources.
Accordingly, in a first step one can use a scaling factor for the neutrino flux upper limits of $1.4\approx1/0.7$ to account for 
the contributions of the blazars that are not in our sample.
In the scenario that high-energy $\gamma$-rays from blazars are ``dissipated'', i.e. isotropized in EBL-induced $\gamma \gamma$-cascades due to intergalactic magnetic fields \citep{Aharonian1994}, the fraction of neutrinos emitted from the sources not resolved in $\gamma$-rays could be higher. 
A simple estimation based on numbers from \citet{Ajello2015} shows, however, that even then the scaling factor must be less than $ \approx 2.8$. The total EGB has an intensity of 
$11.3 \times 10^{-6} \ \mathrm{photons}/\mathrm{cm}^2 \ \mathrm{s} \ \mathrm{sr}$, of which
$4.1 \times 10^{-6} \ \mathrm{photons}/\mathrm{cm}^2 \ \mathrm{s} \ \mathrm{sr}$ originates from resolved blazars and the other 
$7.2 \times 10^{-6} \ \mathrm{photons}/\mathrm{cm}^2 \ \mathrm{s} \ \mathrm{sr}$ from the IGRB (isotropic $\gamma$-ray background). 
If we assume that the entire contribution to the IGRB stems from EBL induced $\gamma \gamma$-cascades from $\gamma$-rays emitted by blazars, i.e. unresolvable isotropized $\gamma$-rays, the resulting ratio between the total emission from blazars and the emission from resolved blazars would be $11.3/4.1 \approx 2.8$. Since the intensity of the IGRB can be well explained by contributions from unresolved - but in principle resolvable - blazars, from star-forming galaxies, and from radio galaxies \citep{Ajello2015}, we deem this maximum cascade emission scenario unlikely and use the factor of $1.43$ throughout this work.

\section{Supplementary figures}
\label{appendix:supplementary}

\begin{figure*}
	\centering
	\begin{tabular}[b]{@{}p{0.45\textwidth}@{}}
		\centering\includegraphics[width=.45\textwidth]{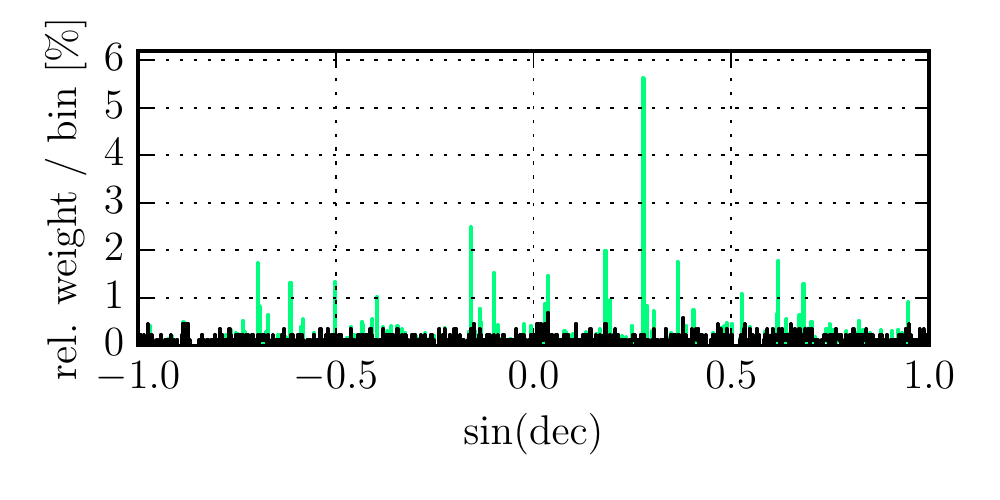} \\
		\centering\small (a) All 2LAC Blazars
	\end{tabular}%
	\quad
	\begin{tabular}[b]{@{}p{0.45\textwidth}@{}}
		\centering\includegraphics[width=.45\textwidth]{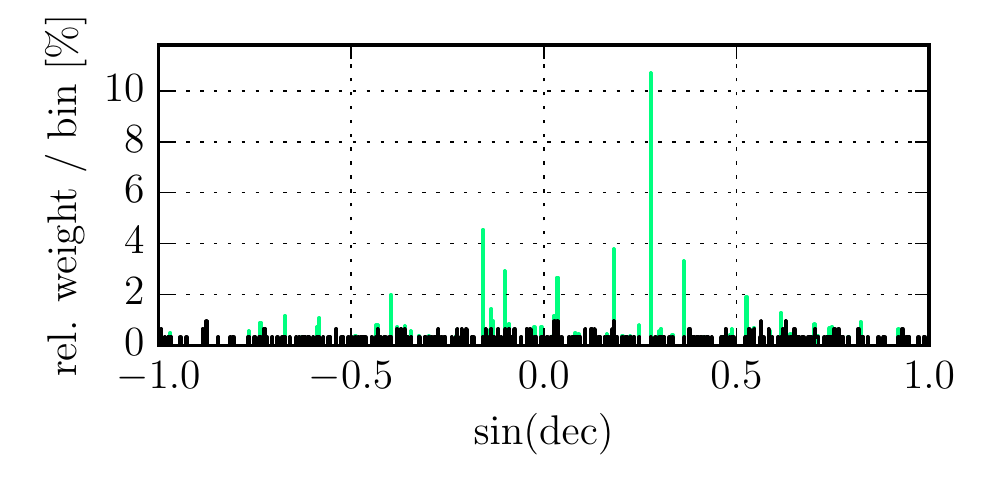} \\
		\centering\small (b) FSRQs
	\end{tabular} \\
	\begin{tabular}[b]{@{}p{0.45\textwidth}@{}}
		\centering\includegraphics[width=.45\textwidth]{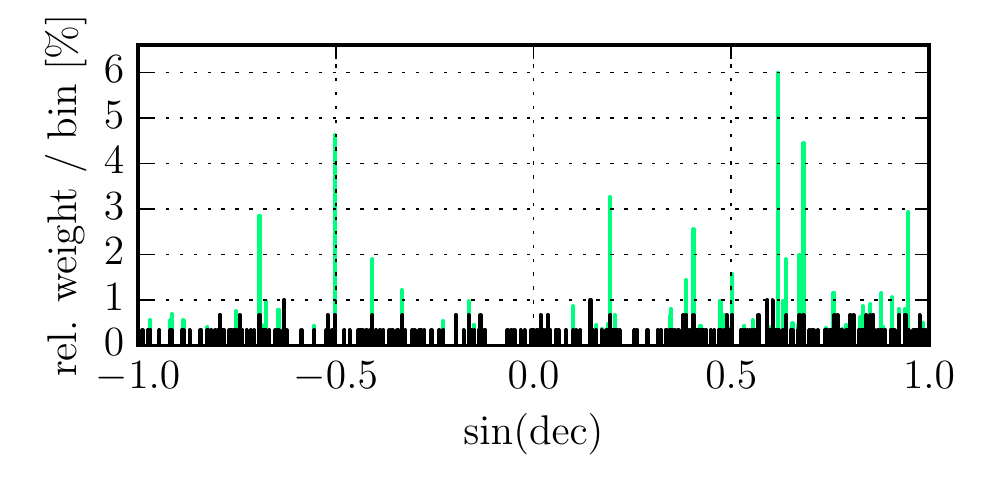} \\
		\centering\small (c) ISP/HSPs
	\end{tabular}%
	\quad
	\begin{tabular}[b]{@{}p{0.45\textwidth}@{}}
		\centering\includegraphics[width=.45\textwidth]{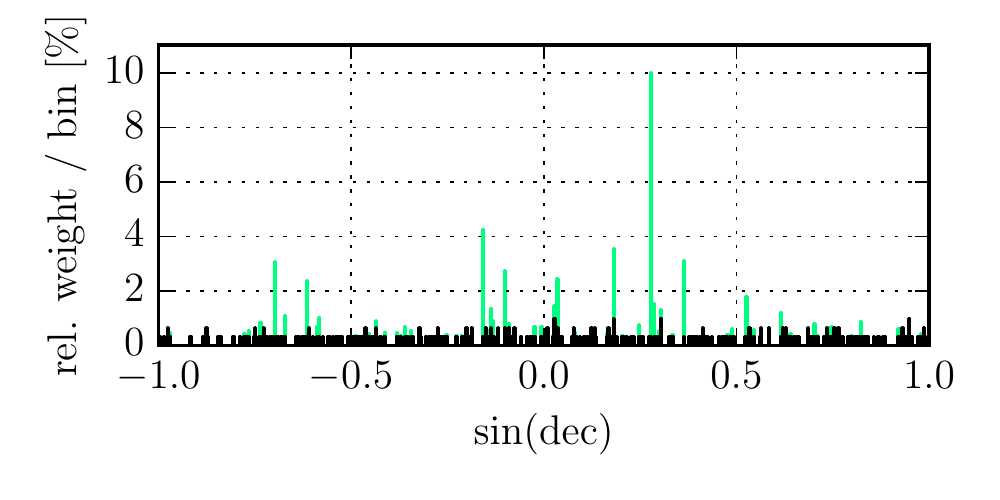} \\
		\centering\small (d) LSPs
	\end{tabular}%
	\begin{tabular}[b]{@{}p{0.45\textwidth}@{}}
		\centering\includegraphics[width=.45\textwidth]{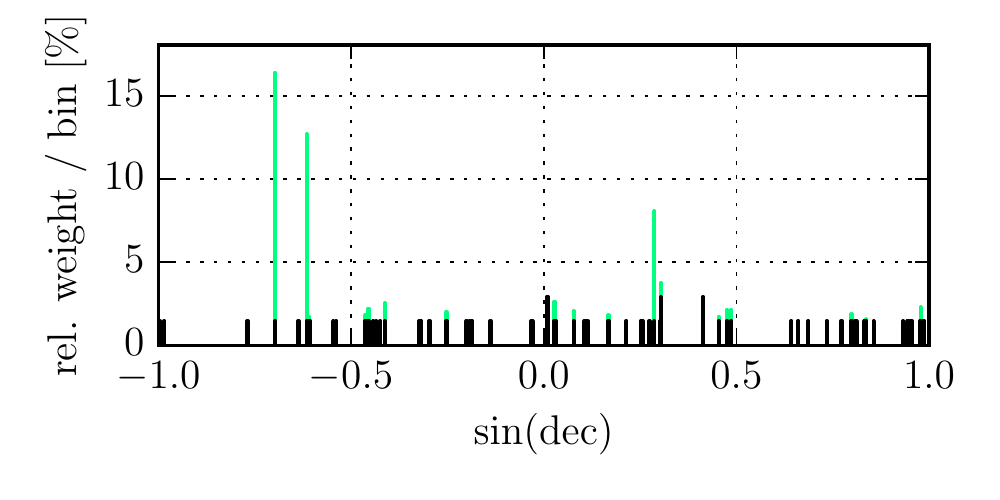} \\
		\centering\small (e) LSP-BL Lacs
	\end{tabular}%
	\caption{Relative contribution to the total sum of all source weights for a given declination bin (in percent). The $\gamma$-weighting scheme is shown in green and the equal weighting scheme is shown in black. The binning is chosen such that at most 5 sources fall into a bin for the largest sample.}
	\label{fig:histogrammed_weights}
\end{figure*}

\begin{figure*}
	\centering
	\begin{tabular}[b]{@{}p{0.45\textwidth}@{}}
		\centering\includegraphics[width=.45\textwidth]{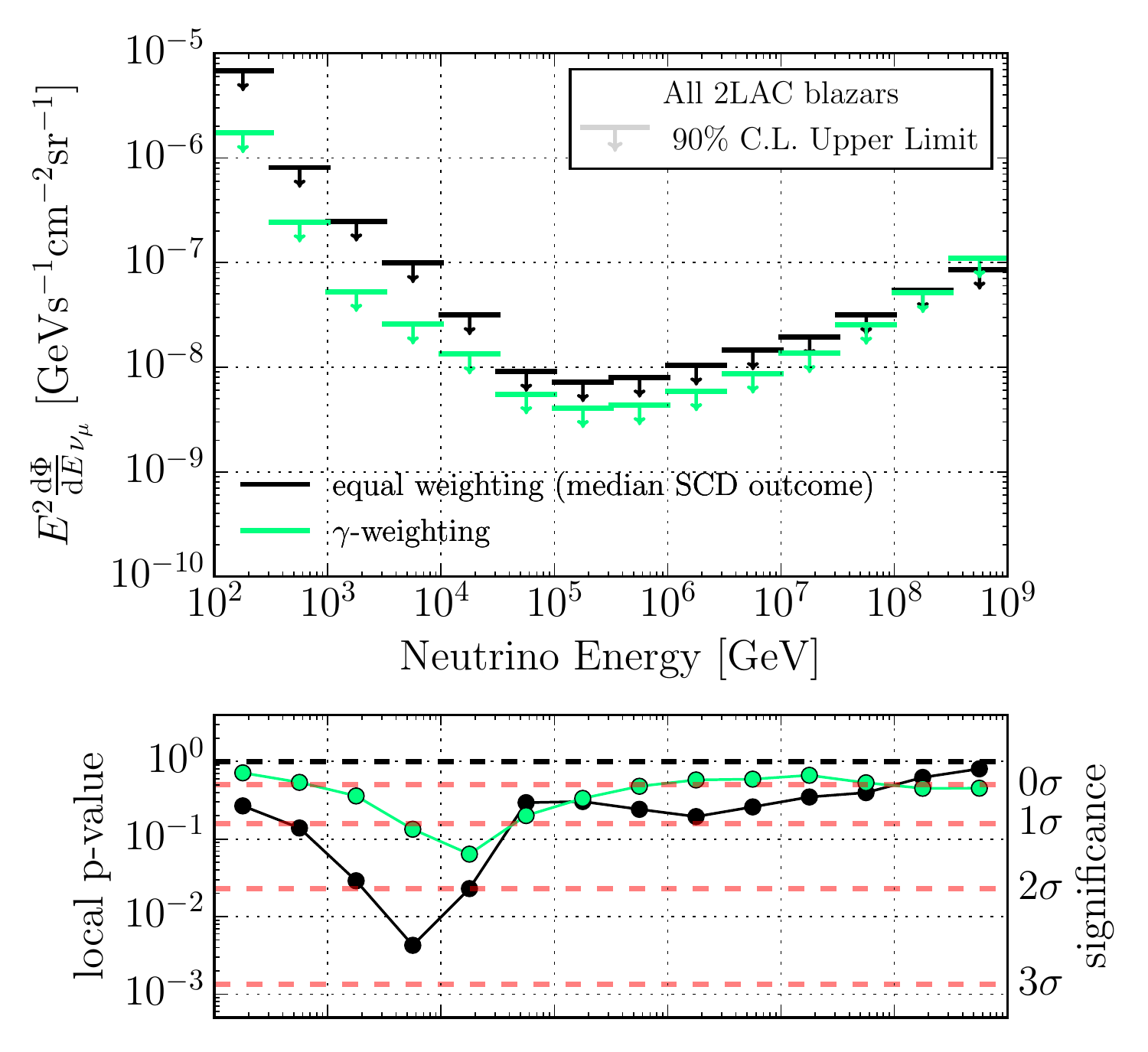} \\
		\centering\small (a) All 2LAC Blazars
	\end{tabular}%
	\quad
	\begin{tabular}[b]{@{}p{0.45\textwidth}@{}}
		\centering\includegraphics[width=.45\textwidth]{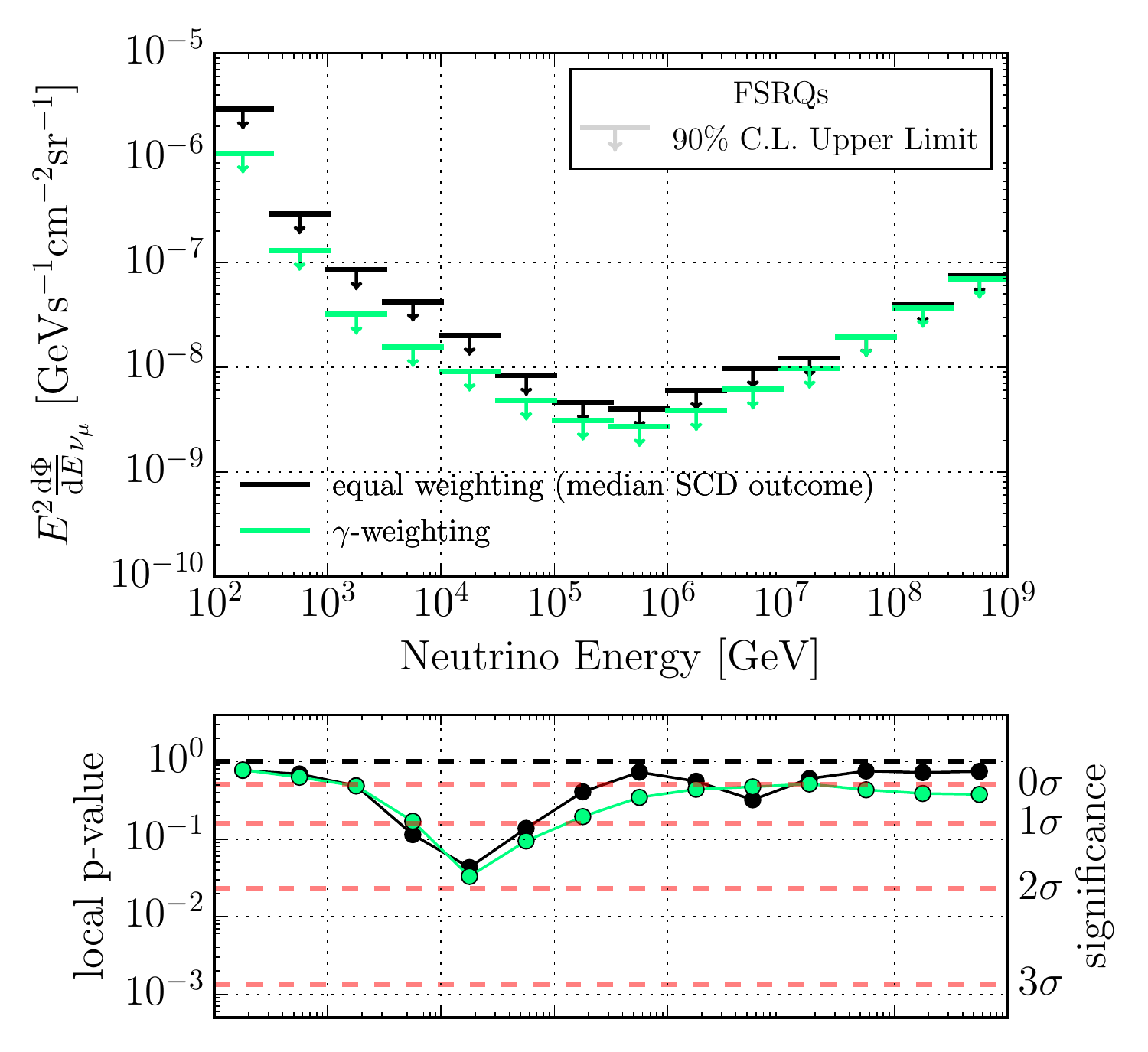} \\
		\centering\small (b) FSRQs
	\end{tabular} \\
	\begin{tabular}[b]{@{}p{0.45\textwidth}@{}}
		\centering\includegraphics[width=.45\textwidth]{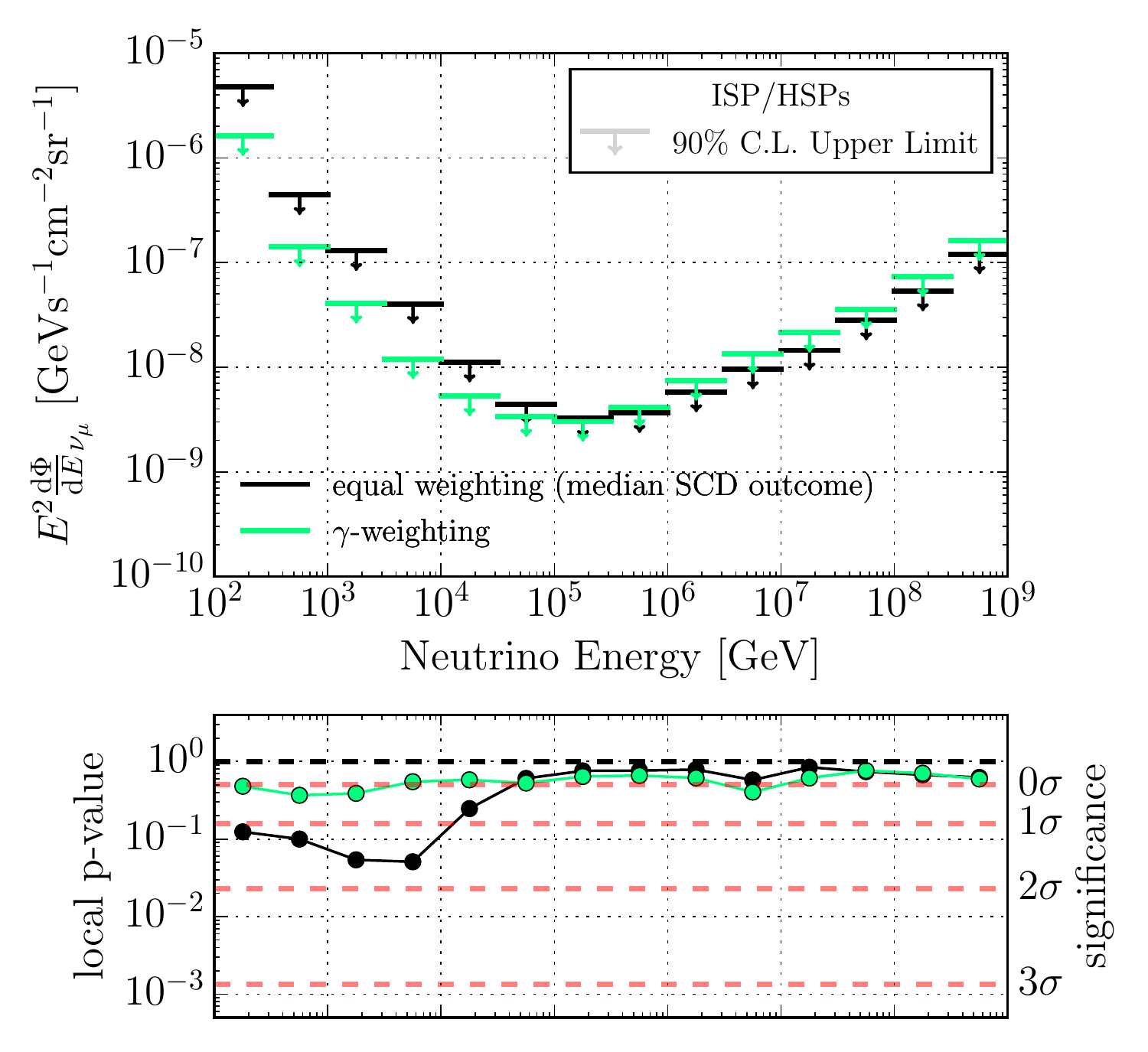} \\
		\centering\small (c) ISP/HSPs
	\end{tabular}%
	\quad
	\begin{tabular}[b]{@{}p{0.45\textwidth}@{}}
		\centering\includegraphics[width=.45\textwidth]{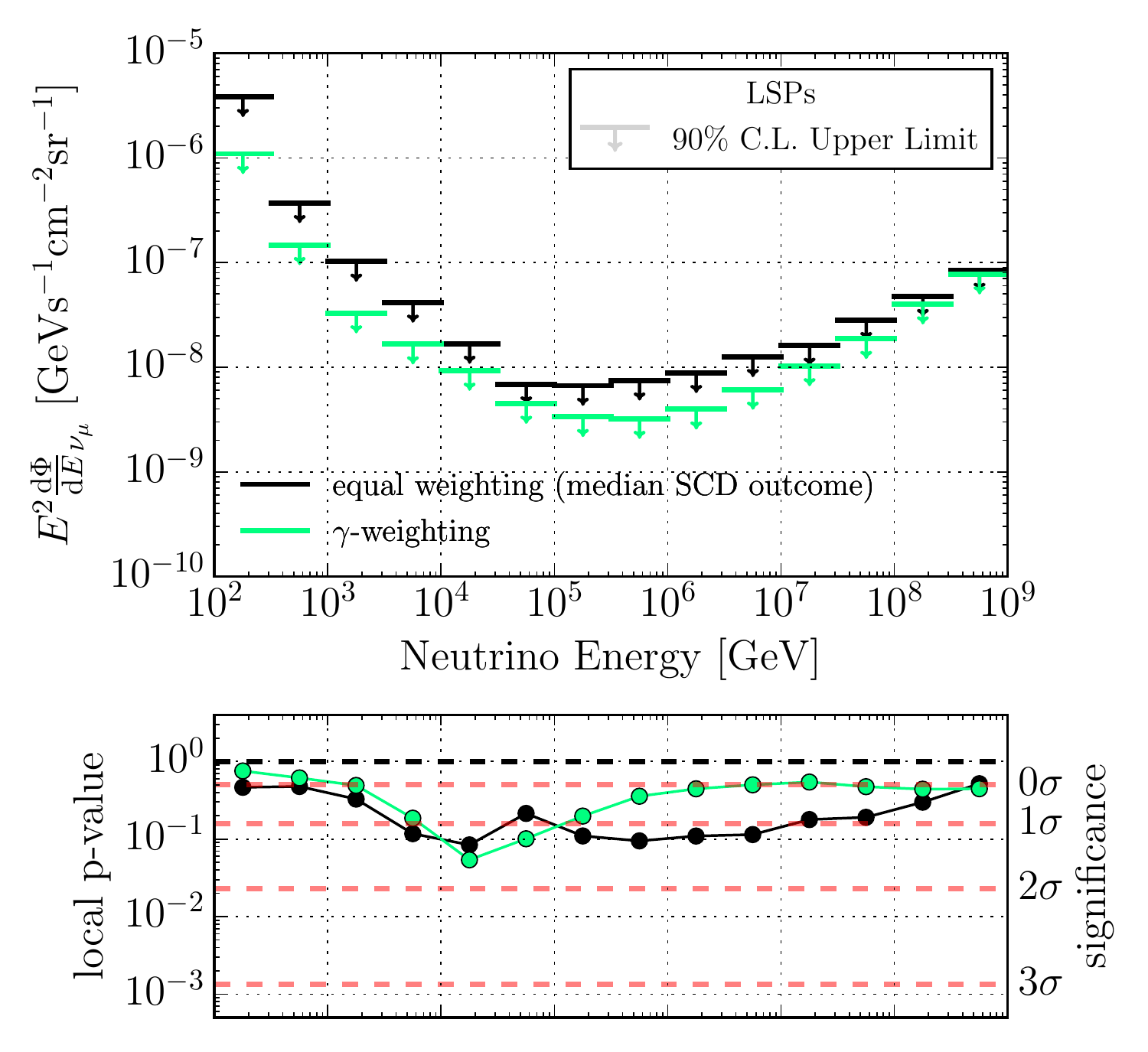} \\
		\centering\small (d) LSPs
	\end{tabular}%
	\begin{tabular}[b]{@{}p{0.45\textwidth}@{}}
		\centering\includegraphics[width=.45\textwidth]{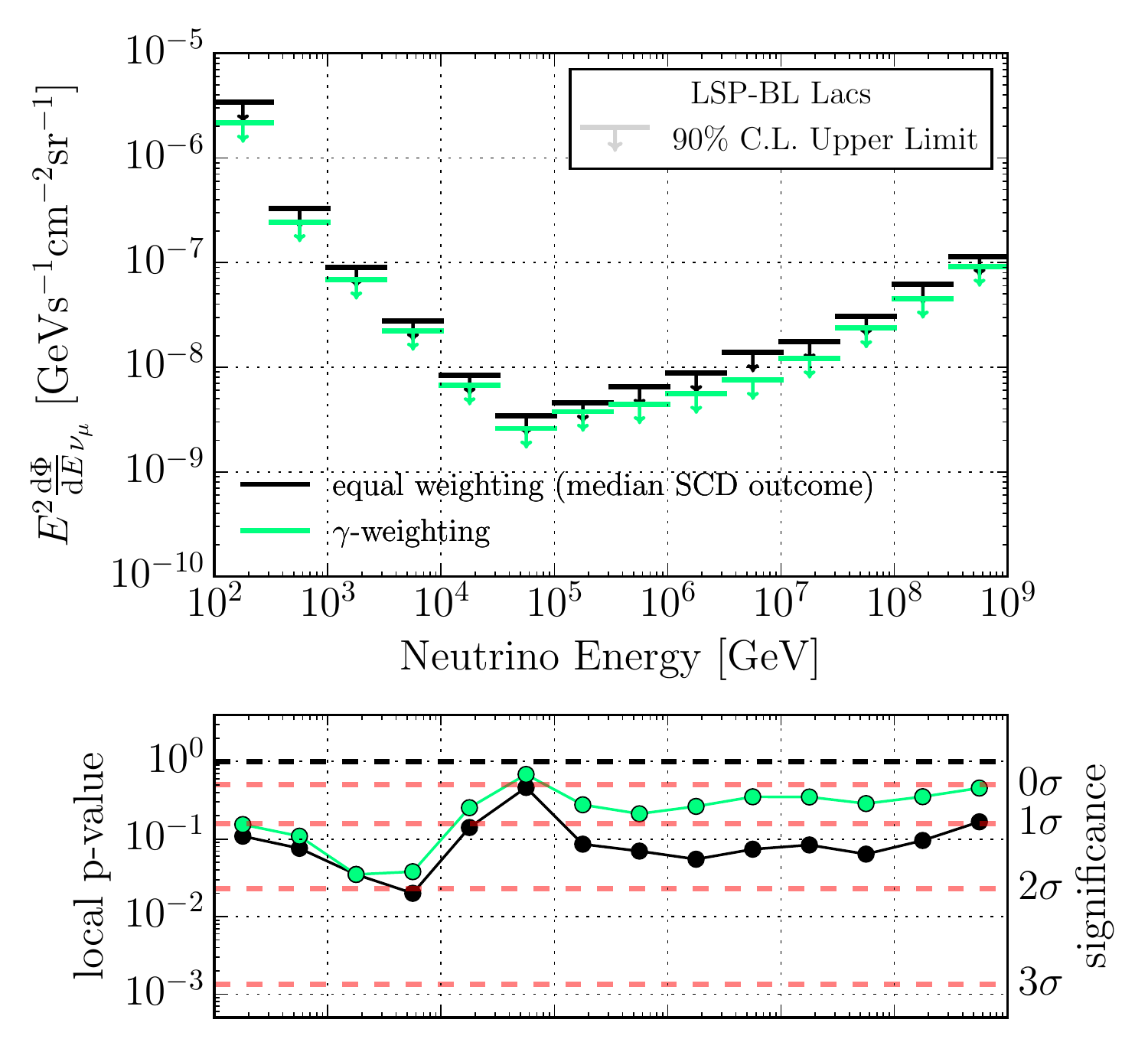} \\
		\centering\small (e) LSP-BL Lacs
	\end{tabular}%
	\caption{$90 \% $ C.L. differential upper limits on the diffuse ($\nu_\mu+\overline{\nu}_\mu$)-flux and corresponding local p-values for the different blazar populations. The equal-weighting upper limits represent the median SCD sampling outcome.}
	\label{fig:differential_limits_and_pvalues}
\end{figure*}






\newpage

\bibliography{manuscript}

\begin{thebibliography}{}
\expandafter\ifx\csname natexlab\endcsname\relax\def\natexlab#1{#1}\fi

\bibitem[{Aartsen {et~al.}(2013{\natexlab{a}})}]{Aartsen2013}
Aartsen, M.~G., {et~al.} 2013{\natexlab{a}}, Science, 342, 1242856

\bibitem[{Aartsen {et~al.}(2013{\natexlab{b}})}]{Aartsen2013a}
---. 2013{\natexlab{b}}, The Astrophysical Journal, 779, 132

\bibitem[{Aartsen {et~al.}(2014{\natexlab{a}})}]{Aartsen2014a}
---. 2014{\natexlab{a}}, Physical Review, D89, 062007

\bibitem[{Aartsen {et~al.}(2014{\natexlab{b}})}]{Aartsen2014}
---. 2014{\natexlab{b}}, The Astrophysical Journal, 796, 109

\bibitem[{Aartsen {et~al.}(2015{\natexlab{a}})}]{Aartsen2015}
---. 2015{\natexlab{a}}, Physical Review, D91, 022001

\bibitem[{Aartsen {et~al.}(2015{\natexlab{b}})}]{Aartsen2015a}
---. 2015{\natexlab{b}}, The Astrophysical Journal, 809, 98

\bibitem[{Aartsen {et~al.}(2015{\natexlab{c}})}]{Aartsen2015b}
---. 2015{\natexlab{c}}, arXiv:1510.05222

\bibitem[{Aartsen {et~al.}(2016{\natexlab{a}})}]{Aartsen2016b}
---. 2016{\natexlab{a}}, accepted by Physical Review Letters, arXiv:1607.05886

\bibitem[{Aartsen {et~al.}(2016{\natexlab{b}})}]{Aartsen2016}
---. 2016{\natexlab{b}}, accepted by Astrophysical Journal, arXiv:1607.08006

\bibitem[{Abbasi {et~al.}(2009)}]{Abbasi2009}
Abbasi, R., {et~al.} 2009, Nuclear Instruments and Methods in Physics Research,
  A601, 294

\bibitem[{Abbasi {et~al.}(2011)}]{Abbasi2011}
---. 2011, Physical Review, D84, 082001

\bibitem[{Abdo {et~al.}(2010{\natexlab{a}})Abdo, Ackermann, Ajello, Axelsson,
  Baldini, {et~al.}}]{Abdo2010}
Abdo, A., Ackermann, M., Ajello, M., {et~al.} 2010{\natexlab{a}}, The
  Astrophysical Journal, 716, 30

\bibitem[{Abdo {et~al.}(2010{\natexlab{b}})Abdo, Ackermann, Ajello,
  {et~al.}}]{Abdo2010b}
---. 2010{\natexlab{b}}, The Astrophysical Journal Supplement Series, 188, 405

\bibitem[{Abdo {et~al.}(2010{\natexlab{c}})Abdo, Ackermann, Ajello, Antolini,
  Baldini, Ballet, Barbiellini, Bastieri, Baughman, Bechtol, \&
  et~al.}]{Abdo2010a}
---. 2010{\natexlab{c}}, The Astrophysical Journal, 720, 435

\bibitem[{Ackermann {et~al.}(2011)Ackermann, Ajello, Allafort, Antolini,
  Atwood, Axelsson, Baldini, Ballet, Barbiellini, Bastieri, \&
  et~al.}]{Ackermann2011}
Ackermann, M., Ajello, M., Allafort, A., {et~al.} 2011, The Astrophysical
  Journal, 743, 171

\bibitem[{Ackermann {et~al.}(2013)Ackermann, Ajello, Allafort, Atwood, Baldini,
  Ballet, Barbiellini, Bastieri, Bechtol, Belfiore, \& et~al.}]{Ackermann2013}
---. 2013, The Astrophysical Journal Supplement Series, 209, 34

\bibitem[{Ackermann {et~al.}(2015)Ackermann, Ajello, Atwood, Baldini, Ballet,
  Barbiellini, Bastieri, Gonzalez, Bellazzini, Bissaldi,
  {et~al.}}]{Ackermann2015}
Ackermann, M., Ajello, M., Atwood, W.~B., {et~al.} 2015, The Astrophysical
  Journal, 810, 14

\bibitem[{Ackermann {et~al.}(2016)Ackermann, Ajello, Atwood, Baldini, Ballet,
  Barbiellini, Bastieri, Gonzalez, Bellazzini, Bissaldi, \&
  et~al.}]{Ackermann2016}
---. 2016, The Astrophysical Journal Supplement Series, 222, 5

\bibitem[{Aharonian {et~al.}(1994)Aharonian, Coppi, \& Volk}]{Aharonian1994}
Aharonian, F.~A., Coppi, P.~S., \& Volk, H.~J. 1994, The Astrophysical Journal,
  423, L5

\bibitem[{Ahlers {et~al.}(2016)Ahlers, Bai, Barger, \& Lu}]{Ahlers2015}
Ahlers, M., Bai, Y., Barger, V., \& Lu, R. 2016, Physical Review, 93, 013009

\bibitem[{Ajello {et~al.}(2015)Ajello, Gasparrini, S{\'a}nchez-Conde,
  Zaharijas, Gustafsson, {et~al.}}]{Ajello2015}
Ajello, M., Gasparrini, D., S{\'a}nchez-Conde, M., {et~al.} 2015, The
  Astrophysical Journal, 800, L27

\bibitem[{Ajello {et~al.}(2014)Ajello, Romani, Gasparrini, Shaw, Bolmer,
  {et~al.}}]{Ajello2014}
Ajello, M., Romani, R., Gasparrini, D., {et~al.} 2014, The Astrophysical
  Journal, 780, 73

\bibitem[{Ajello {et~al.}(2012)Ajello, Shaw, Romani, Dermer, Costamante,
  {et~al.}}]{Ajello2012}
Ajello, M., Shaw, M., Romani, R., {et~al.} 2012, The Astrophysical Journal,
  751, 108

\bibitem[{Atoyan \& Dermer(2001)}]{Atoyan2001}
Atoyan, A., \& Dermer, C. 2001, Physical Review Letters, 87, 221102

\bibitem[{Barlow(1990)}]{Barlow1990}
Barlow, R. 1990, Nuclear Instruments and Methods in Physics Research, A297, 496

\bibitem[{Becker {et~al.}(2005)Becker, Biermann, \& Rhode}]{Becker2005}
Becker, J., Biermann, P., \& Rhode, W. 2005, Astroparticle Physics, 23, 355

\bibitem[{B{\"o}ttcher \& Dermer(2002)}]{Bottcher2002}
B{\"o}ttcher, M., \& Dermer, C. 2002, The Astrophysical Journal, 564, 86

\bibitem[{B{\"o}ttcher {et~al.}(2013)B{\"o}ttcher, Reimer, Sweeney, \&
  Prakash}]{Boettcher2013}
B{\"o}ttcher, M., Reimer, A., Sweeney, K., \& Prakash, A. 2013, The
  Astrophysical Journal, 768, 54

\bibitem[{Bugaev {et~al.}(2004)Bugaev, Montaruli, Shlepin, \&
  Sokalski}]{Bugaev2004}
Bugaev, E., Montaruli, T., Shlepin, Y., \& Sokalski, I. 2004, Astroparticle
  Physics, 21, 491

\bibitem[{Cavaliere \& D'Elia(2002)}]{Cavaliere2002}
Cavaliere, A., \& D'Elia, V. 2002, The Astrophysical Journal, 571, 226

\bibitem[{Dom{\'i}nguez {et~al.}(2013)Dom{\'i}nguez, Finke, Prada, Primack,
  Kitaura, {et~al.}}]{Dominguez2013}
Dom{\'i}nguez, A., Finke, J., Prada, F., {et~al.} 2013, The Astrophysical
  Journal, 770, 77

\bibitem[{Falcke \& Biermann(1995)}]{Falcke1995}
Falcke, H., \& Biermann, P. 1995, Astronomy and Astrophysics, 293, 665

\bibitem[{Fossati {et~al.}(1998)Fossati, Maraschi, Celotti, Comastri, \&
  Ghisellini}]{Fossati1998}
Fossati, G., Maraschi, L., Celotti, A., Comastri, A., \& Ghisellini, G. 1998,
  Monthly Notices of the Royal Astronomical Society, 299, 433

\bibitem[{Georgakakis {et~al.}(2008)Georgakakis, Nandra, Laird, Aird, \&
  Trichas}]{Georgakakis2008a}
Georgakakis, A., Nandra, K., Laird, E.~S., Aird, J., \& Trichas, M. 2008,
  Monthly Notices of the Royal Astronomical Society, 388, 1205

\bibitem[{Giommi {et~al.}(2012)Giommi, Padovani, Polenta, Turriziani, D'Elia,
  {et~al.}}]{Giommi2012}
Giommi, P., Padovani, P., Polenta, G., {et~al.} 2012, Monthly Notices of the
  Royal Astronomical Society, 420, 2899

\bibitem[{Halzen \& Zas(1997)}]{Halzen1997}
Halzen, F., \& Zas, E. 1997, The Astrophysical Journal, 488, 669

\bibitem[{Hill \& Rawlins(2003)}]{Hill2003}
Hill, G., \& Rawlins, K. 2003, Astroparticle Physics, 19, 393

\bibitem[{Hopkins {et~al.}(2003)Hopkins, Afonso, Chan, Cram, Georgakakis, \&
  Mobasher}]{Hopkins2003}
Hopkins, A.~M., Afonso, J., Chan, B., {et~al.} 2003, The Astronomical Journal,
  125, 465

\bibitem[{Kadler {et~al.}(2016)Kadler, Krau{\ss}, Mannheim, Ojha, M{\"u}ller,
  Schulz, Anton, Baumgartner, Beuchert, Buson, {et~al.}}]{Kadler2016}
Kadler, M., Krau{\ss}, F., Mannheim, K., {et~al.} 2016, Nature Physics

\bibitem[{Mannheim(1995)}]{Mannheim1995}
Mannheim, K. 1995, Astroparticle Physics, 3, 295

\bibitem[{Meyer {et~al.}(2011)Meyer, Fossati, Georganopoulos, \&
  Lister}]{Meyer2011}
Meyer, E.~T., Fossati, G., Georganopoulos, M., \& Lister, M.~L. 2011, The
  Astrophysical Journal, 740, 98

\bibitem[{M\"ucke {et~al.}(2003)M\"ucke, Protheroe, Engel, Rachen, \&
  Stanev}]{Muecke2003}
M\"ucke, A., Protheroe, R., Engel, R., Rachen, J., \& Stanev, T. 2003,
  Astroparticle Physics, 18, 593

\bibitem[{Murase {et~al.}(2015)Murase, Guetta, \& Ahlers}]{Murase2015}
Murase, K., Guetta, D., \& Ahlers, M. 2015, Physical Review Letters, 116,
  071101

\bibitem[{Murase {et~al.}(2014)Murase, Inoue, \& Dermer}]{Murase2014}
Murase, K., Inoue, Y., \& Dermer, C. 2014, Physical Review, D90, 023007

\bibitem[{Neronov \& Semikoz(2016)}]{Neronov2016}
Neronov, A., \& Semikoz, D. 2016, Astroparticle Physics, 75, 60–63

\bibitem[{Neunhoffer(2006)}]{Neunhoffer2006}
Neunhoffer, T. 2006, Astroparticle Physics, 25, 220

\bibitem[{Nolan {et~al.}(2012)Nolan, Abdo, Ackermann, Ajello, Allafort,
  {et~al.}}]{Nolan2012}
Nolan, P.~L., Abdo, A.~A., Ackermann, M., {et~al.} 2012, The Astrophysical
  Journal Supplement Series, 199, 31

\bibitem[{Olive(2014)}]{Olive2014}
Olive, K. 2014, Chinese Physics C, 38, 090001

\bibitem[{Padovani \& Giommi(1995)}]{Padovani1995}
Padovani, P., \& Giommi, P. 1995, The Astrophysical Journal, 444, 567

\bibitem[{Padovani {et~al.}(2015)Padovani, Petropoulou, Giommi, \&
  Resconi}]{Padovani2015}
Padovani, P., Petropoulou, M., Giommi, P., \& Resconi, E. 2015, Monthly Notices
  of the Royal Astronomical Society, 452, 1877

\bibitem[{Padovani {et~al.}(2016)Padovani, Resconi, Giommi, Arsioli, \&
  Chang}]{Padovani2016}
Padovani, P., Resconi, E., Giommi, P., Arsioli, B., \& Chang, Y.~L. 2016,
  Monthly Notices of the Royal Astronomical Society, 457, 3582

\bibitem[{{Planck Collaboration}(2015)}]{Planck2015}
{Planck Collaboration}. 2015, accepted by Astronomy \& Astrophysics,
  arXiv:1502.01589

\bibitem[{Protheroe(1997)}]{Protheroe1997}
Protheroe, R. 1997, ASP Conference Series, 121, 585

\bibitem[{Read(2000)}]{Read2000}
Read, A. 2000

\bibitem[{Schuster {et~al.}(2002)Schuster, Pohl, \&
  Schlickeiser}]{Schuster2002}
Schuster, C., Pohl, M., \& Schlickeiser, R. 2002, Astronomy and Astrophysics,
  382, 829

\bibitem[{Stecker {et~al.}(1991)Stecker, Done, Salamon, \&
  Sommers}]{Stecker1991}
Stecker, F.~W., Done, C., Salamon, M.~H., \& Sommers, P. 1991, Physical Review
  Letters, 66, 2697, [Erratum: Phys. Rev. Lett.69,2738(1992)]

\bibitem[{Stickel {et~al.}(1991)Stickel, Fried, Kuehr, Padovani, \&
  Urry}]{Stickel1991}
Stickel, M., Fried, J.~W., Kuehr, H., Padovani, P., \& Urry, C.~M. 1991, The
  Astrophysical Journal, 374, 431

\bibitem[{Tavecchio \& Ghisellini(2015)}]{Tavecchio2015}
Tavecchio, F., \& Ghisellini, G. 2015, Monthly Notices of the Royal
  Astronomical Society, 451, 1502

\bibitem[{Tavecchio {et~al.}(2014)Tavecchio, Ghisellini, \&
  Guetta}]{Tavecchio2014}
Tavecchio, F., Ghisellini, G., \& Guetta, D. 2014, The Astrophysical Journal,
  793, L18

\bibitem[{Urry \& Padovani(1995)}]{Urry1995}
Urry, C., \& Padovani, P. 1995, Publications of the Astronomical Society of the
  Pacific, 107, 803

\bibitem[{Wang \& Li(2015)}]{Wang2015}
Wang, B., \& Li, Z. 2015, Science China Physics, Mechanics \& Astronomy, 59, 1

\end{thebibliography}

\end{document}